\documentclass[epj]{svjour}

\usepackage{bm}% bold math
\usepackage{latexsym}
\usepackage{graphicx}% graphics
\usepackage{feynmp}     
\def\pp{particle-particle}
\def\sus{susceptibility}
\def\pam{periodic Anderson model}
\def\Sus{Susceptibility}
\def\suss{susceptibilities}

\def\HF{Heavy Fermion}
\def\se{self-energy}
\def\etal{{\em et.~al.}}
\def\NFL{non-Fermi liquid}
\def\redu{reducible}
\def\irr{ir\redu}
\def\irr{ir\redu}
\def\irrep{irreducible representations}
\def\gf{Green function}
\def\Gf{\gf}
\def\GF{Green Function}
\def\onep{one-particle}
\def\twop{two-particle}
\def\tc{two-channel}
\def\oddf{odd-frequency}
\def\tcpam{\tc\ \pam}
\def\hamil{Hamiltonian}
\def\IV{intermediate valence}
\def\ket#1{|#1\rangle}
\def\bra#1{\langle  #1 |}
\def\w{\omega}
\def\e{\varepsilon}

\def\as{\alpha\sigma}
\def\eas{\e_{\alpha\sigma}}
\def\aps{\alpha'\sigma}
\def\asp{\alpha\sigma'}
\def\apsp{\alpha'\sigma'}
\def\k{\vec{k}}
\def\Q{\vec{Q}}
\def\kas{\k\as}
\def\komma{\; ,\;}
\def\punkt{\; .\;}
\def\corr#1#2#3{\ll \! #1 | #2 \!\gg (#3)}
\def\corrz#1#2{\corr{#1}{#2}{z}}
\def\dps{\displaystyle}
\def\iw#1{i\omega_{#1}}

\def\iwnn{\iw{n'}}
\def\iwm{i\w_m}
\def\iwn{i\w_n}
\def\iw#1{i\w_{#1}}
\def\ivn{i\nu_n}
\def\q{\vec{q}}
\def\kp{\vec{k'}}
\def\ul#1{\underline{#1}}
\def\mat#1{\ul{\ul{#1}}}
\def\myCint{\frac{1}{\tilde Z_{eff}}\oint_{{\cal C}}}
\def\expect#1{\langle #1 \rangle}
\def\kk{\kp}
\def\onepart#1#2#3{ G_{#1, #2}(#3)}
\def\onepartz#1#2{\onepart{#1}{#2}{z}}
\def\invmat#1{\left[ #1 \right]^{-1}}
\def\spur#1{\mbox{Tr}\left[#1\right]}
\def\non{\nonumber \\ }

\begin{document}

\begin{fmffile}{fmf-odd-f-sc}
%%---------------------------------
%% macros for the two particle
%% pair propagator
%%--------------------------------
%%
%%------------------------------
%%  unfilled vextex
%%--------------------------------
\def\fmfunblob#1#2{%%
\fmfv{decor.shape=circle,decor.filled=empty,decor.size=#1}{#2}%%
}
%%------------------------------
%%  filled vextex
%%--------------------------------
\def\fmfmydot#1#2{%%
\fmfv{decor.shape=circle,decor.filled=full,decor.size=#1}{#2}%%
}
%%----------------------------------------
%%
%%------------------------------------
\def\fmfPairProp#1#2#3#4{
\Large
\begin{fmfchar*}(110,40)
 \fmfleft{cin} 
  \fmfmydot{0.05w}{cin}
 \fmf{fermion,left=0.1,label=#1}{cin,h2}  
 \fmf{fermion,right=0.1,label=#2}{cin,h1}
 \fmfrpolyn{empty,l=$I$}{h}{4}
 \fmf{fermion,left=0.1,label=#3}{h3,cout}  
 \fmf{fermion,right=0.1,label=#4}{h4,cout}
 \fmfright{cout} 
 \fmfmydot{0.05w}{cout}
\end{fmfchar*}
}
\unitlength=1mm
%%------------------------------
%%  free propagator
%% 
%%  #1 propagator type lower
%%  #2 proptype  higher
%%
%%  #3 label low
%%  #4 label high
%%--------------------------------
\def\fmfFreePropBare#1#2#3#4{
\begin{fmfchar*}(20,20)
  \fmfpen{thick}
 \fmfleft{cin1,cin2}
 \fmfright{cout1,cout2}
 \fmf{#1,label=#3}{cin1,cout1}
 \fmf{#2,label=#4}{cin2,cout2}
\end{fmfchar*}
}
%%------------------------------
%%  free fermionic propagator
%% 
%%  #1 label low
%%  #2 label high
%%--------------------------------
\def\fmfFreeBandProp#1#2{\fmfFreePropBare{fermion}{fermion}{#1}{#2}}
%%------------------------------
%%  free exchange propagator
%% 
%%  #1 propagator type lower
%%  #2 label 
%%--------------------------------
\def\fmfFreeExPropBare#1#2{
\begin{fmfchar*}(20,20)
 \fmfpen{thick}
 \fmfleft{cin1,cin2}
 \fmfright{cout1,cout2}
 \fmf{plain}{cin1,v}
 \fmf{plain}{cin2,v}
 \fmf{#1,label=#2}{v,cout2}
 \fmf{#1,label=#2}{v,cout1}
\end{fmfchar*}
}
%%------------------------------
%%  two partice propagator for two different
%% particels
%% 
%%  #1 particle type 1  low
%%  #2 particle type 2  high
%%
%%  #3 label in the center
%%
%%  #4 label low  in
%%  #5 label high in
%%  #6 label low  out
%%  #7 label high out
%%--------------------------------
\def\fmfTwoPartProp#1#2#3#4#5#6#7{
{%%\large
\begin{fmfchar*}(40,20)
 \fmfpen{thick}
 \fmfleft{cin,fin}
 \fmfright{cout,fout}
 \fmf{#1,label=#4}{cin,h1}
 \fmf{#2,label=#5}{fin,h2}
 \fmfrpolyn{empty,label=#3}{h}{4}
 \fmf{#1,label=#6}{h4,cout}
 \fmf{#2,label=#7}{h3,fout}
\end{fmfchar*}
}
}
%%------------------------------
%%  Dyson equation for two identical
%% particles
%% 
%%  #1 label in the center of Gamma
%%  #2 label in the center of Two_part
%%
%%  #3 label low  in
%%  #4 label high in
%%
%%  #5 label low  connect
%%  #6 label high connect
%%
%%  #7 label low  out
%%  #8 label high out
%%--------------------------------
\def\fmfTwoPartDyson#1#2#3#4#5#6#7#8{
{%%\large
\begin{fmfchar*}(70,30)
 \fmfpen{thick}
 \fmfleft{cin,fin}
 \fmfright{cout,fout}
 \fmf{fermion,label=#3}{cin,G1}
 \fmf{fermion,label=#4}{fin,G2}
 \fmfrpolyn{empty,label=#1}{G}{4}
 \fmf{fermion,label=#5,label.side=right}{G4,N1}
 \fmf{fermion,label=#6,label.side=left}{G3,N2}
 \fmfrpolyn{empty,label=#2}{N}{4}
 \fmf{fermion,label=#7}{N4,cout}
 \fmf{fermion,label=#8}{N3,fout}
\end{fmfchar*}
}
}
%%------------------------------
%%  two particle band  propagator
%%
%%  #1 label in the center
%%
%%  #2 label low  in
%%  #3 label high in
%%  #4 label low  out
%%  #5 label high out
%%--------------------------------
\def\fmfTwoPartBandProp#1#2#3#4#5{\fmfTwoPartProp{fermion}{fermion}
{#1}{#2}{#3}{#4}{#5}
}
%%------------------------------
%%  two partice propagator for two different
%% particels
%% 
%%  #1 particle type 1  low
%%  #2 particle type 2  high
%%
%%  #3 label in the center
%%
%%  #4 label low  in
%%  #5 label high in
%%  #6 label low  out
%%  #7 label high out
%%--------------------------------
\def\fmfTwoPartDirProp#1#2#3#4#5#6#7{
{%%\large
\begin{fmfchar*}(45,15)
 \fmfpen{thick}
 \fmfleft{cin,fin}
  \fmftopn{t}{4}
  \fmfbottomn{b}{4}
 \fmfright{cout,fout}
 \fmf{#1,label=#4}{cin,b2}
 \fmf{#2,label=#5}{fin,t2}
 \fmf{plain}{b2,b3,t3,t2,b2}
 \fmf{#1,label=#6}{b3,cout}
 \fmf{#2,label=#7}{t3,fout}
\end{fmfchar*}
}
}
\def\fmfTwoPartExProp#1#2#3#4#5#6#7{
{%%\large
\begin{fmfchar*}(45,15)
 \fmfpen{thick}
 \fmfleft{cin,fin}
  \fmftopn{t}{4}
  \fmfbottomn{b}{4}
 \fmfright{cout,fout}
 \fmf{#1,label=#4,label.side=left}{cin,t2}
 \fmf{#2,label=#5,label.side=right}{fin,b2}
 \fmf{plain}{b2,b3,t3,t2,b2}
 \fmf{#1,label=#6}{b3,cout}
 \fmf{#2,label=#7}{t3,fout}
\end{fmfchar*}
}
}
%%
%%--------------------
%%  EXAMPLE
%%---------------------------
%%\begin{picture}(150,100)
%%\put(20,35){\fmfPairProp{$\sigma\alpha$}{$\sigma'\alpha'$}
%%{$\sigma''\alpha''$}{$\sigma'''\alpha'''$}
%%}
%%\end{picture}

\title{Composite Spin-Triplet Superconductivity 
  in an $SU(2)\otimes SU(2)$ Symmetric Lattice Model}

%\smallskip
\author{Frithjof B. Anders\inst{1} }
\institute{Institut f\"ur Festk\"orperphysik, Technical University
 Darmstadt, 64289 Darmstadt, Germany
\email{frithjof@fkp.tu-darmstadt.de}
}
%\begin{abstract}
\abstract{
The two-channel Anderson lattice model which has $SU(2)\otimes SU(2)$
symmetry is of relevance to  understanding of the magnetic, quadrupolar and
superconducting phases in U$_{1-x}$Th$_x$Be$_{13}$ or Pr base
skutterudite compounds such as PrFe$_4$P$_{12}$ or PrOs$_4$Sb$_{12}$.
Possible  unconventional superconducting phases   of the model are
explored. They are characterized by a composite order parameter
comprising of a local magnetic or quadrupolar moment and a triplet
conduction electron Cooper-pair. This binding of local degrees of
freedom removes the entropy of the non Fermi-liquid normal state. We
find  superconducting transitions in the intermediate valence regime
which are suppressed in the stable moment regime. The gap function is
non analytic and odd in frequency: a pseudo-gap develops in the
conduction electron density of states which vanishes as $|\omega|$
close to $\omega=0$. In the strong intermediate valent regime, the gap
function acquires an additional $\k$-dependence. 
}
%\end{abstract}

%\pacs{75.20.Hr, 75.30.Mb, 71.27.+a} 

\maketitle

\def\hel{$^3$He}

\section{Introduction}
\label{introduction}

Superconductivity or superfluidity of  \hel\ \cite{WoelfleVollhardt} is
understood to be a thermodynamic 
phase in which  spin $1/2$ Fermions are subject to Cooper-pair
correlations and condense. The
Fermions are either electrons  or
\hel\ atoms. %%Conventional
Superconductivity is  well described by the
Bardeen-Cooper-Schrieffer (BCS) theory \cite{BCS57} when the
fluctuations of the order parameter are small.
The Eliashberg theory of superconductivity \cite{Eliashberg60,AllenMitrovic} is
applicable in superconductors when high and low  energy scales are
well  separated. Contrary to the BCS theory, this theory 
accounts for  the retardation of the  interaction mediating Bosons.
On a large time scale, the electron repulsion is effectively overcome
in standard superconductors  by the electron-phonon interaction. The
small parameter needed for the applicability of the Eliashberg theory is
given  by the ratio of the Debye frequency and the Fermi energy. It is well
known  that Eliashberg theory cannot be used in the \hel\ problem
\cite{WoelfleVollhardt} since no separation of energy scales can be
found here. \hel\ atoms pair in spin triplets and orbital momentum of
$L=1$ in contrast to standard superconductors which usually develop
spin singlets pairing. The rich phase diagram of \hel\ is directly
related to this anisotropic pair-condensation. Pairs with $L=1$ have
nodes in space and, therefore, can  effectively overcome the hard-core
repulsion and gain condensation energy.  

Heavy Fermion  materials \cite{Grewe91} have drawn
much attention since the discovery of
superconductivity in CeCu$_2$Si$_2$ \cite{Steglich79}, which seems to be
characterized by an anisotropic order parameter with a symmetry yet  to be
determined. Another example for an anisotropic order parameter are the
high temperature cuprate superconductors. Their current carrying
subsystem consists of layered copper oxide perovskite structures.
Strong experimental evidence has been compiled for 
an unconventional order parameter  in these materials, with $\Gamma_2$ symmetry  
of the form $\cos (k_x a) -\cos (k_y a) $ or $(k_x^2-k_y^2)$ which is
usually called ``$d$-wave'' superconductivity by many authors.
%, even though the order
%parameter is a scalar and does not have five components as a true
%$d$-wave superconductor.
A superconductor is called {\em conventional} if the order parameter has the
full  symmetry of the Fermi-surface; the order parameter still might
be anisotropic but it transforms in accordance with the trivial
irreducible representation of the  point-group $\Gamma_1$. Otherwise,
the order parameter is called {\em unconventional}.
Recently, the superconductivity of Sr$_2$RuO$_4$ has generated strong
interest due to structural similarities between this compound and the 
{\em High-$T_c$} materials \cite{Maeno94}. It has been suggested that
Sr$_2$RuO$_4$ might be an example of an unconventional  spin triplet
superconductor \cite{SigristRise95,Baskaran96} in contrast to the {\em
  High-$T_c$} materials where a $\Gamma_2$ symmetry is favoured. 
In spite of the
progress in our  understanding of \HF\ materials, 
there has been little  success in  determining the symmetry of their
order parameters in the superconducting phase.
Huth \etal\ recently reported experimental evidence
which indicates an electronically mediated pair-condensation mechanism in
UPd$_2$Al$_3$ \cite{Huth98}. Our theoretical understanding of
unconventional superconductivity is still
far from complete in spite of more than 20 years of research. Theories
were developed along two basic lines: 
either the superconducting phase of 
microscopic models was investigated using, at that time, state of
the art approximations in order to reveal possible pairing mechanisms,
or phenomenological approaches, based on the Ginsburg-Landau theory, were
constructed in order to give  new insight into the possible symmetries of
the order parameter and their impact on the measurable
quantities \cite{SCcat}.
%\footnote{For a 
%review of the basic ideas see Grewe and Steglich
%\cite{Grewe91}). The literature of the last decade can be
%classified by these two rather crued categories.}.

Two channel Anderson and Kondo lattice models
 exhibit non-Fermi liquid
behaviour in the paramagnetic phase driven by unquenched and
fluctuating local degrees of freedom \cite{AndersJarCox97,JarrellPangCox97}.
This phase is characterized by a large residual resistivity and
entropy, and ill defined electronic quasi-particles. Fermi liquid
physics is restored by cooperative ordering or applied magnetic field
or stress \cite{JarrellPangCox97,Anders99}.
In particular, for a non Kramer's doublet crystal field state in U$^{4+}$
or Pr$^{3+}$ ions, a two channel Kondo effect is possible
\cite{Cox87}. The magnetic ``spin'' of the electrons serves as channel
index in this case. 
UBe$_{13}$ and PrFeP$_{12}$ are prominent candidates for a quadrupolar
Kondo lattice description, as their enormous resistivity
($>100\mu\Omega cm$) is removed only by phase transitions
(superconductivity in UBe$_{13}$, antiferroquadrupolar order in
PrFeP$_{12}$). Indeed, it has been shown that commensurate and
incommensurate orbital ordering, as well as ferromagnetism are
possible in this model depending upon coupling strength and
filling \cite{Anders99}. Evidence of a first order transition to an odd
frequency  pairing state in the Kondo limit (near integral valence) 
has been adduced, which  is a singlet in both spin and
channel indices, and no $\q$ value was preferred for the center of mass
momentum  \cite{JarrellPangCox97}.

Our paper is organized as follows. In section \ref{sec:theory}, we
review the two-channel Anderson 
lattice model \cite{AndersJarCox97}.
This model is  rotational
invariant in spin and channel space and, therefore, has $SU(2)\otimes
SU(2)$ symmetry. We use
established standard methods developed by others
\cite{AndersJarCox97,JarrellPangCox97,Anders99,Grewe87,Kim87,MetznerVollhardt89,Kim90,Jarrell95,Pruschke95,Georges96}  
  to obtain an  approximate solution of 
its one- and two-particle properties. 
In section \ref{sec:superconductivity}, all possible spin and channel
symmetry sectors of Cooper-pairs are discussed. Two of the four
sectors require 
odd-frequency  gap functions \cite{Berezinskii74} if we restrict ourselves
to  $\k$-dependence with $\Gamma_1$ symmetry. 
We discuss  the concept of a composite
order parameter in section
\ref{sec:composite-orderparameter}. A connection between the composite
order parameter and the Cooper-pair \sus\ is
made in Sec.\ \ref{sec:composite-pp}. The statements of Section
\ref{sec:superconductivity} are exact and independent of the any
approximation. 
The solution of the homogeneous phase of the
model \cite{AndersJarCox97} is used to obtain the superconducting
transition temperature defined by the divergence of the pair-\sus\ which
will be the subject of Sec. \ref{sec:phase-boundary}. In section
\ref{sec:dmft-sc}, we present the self-consistency equations for the
triplet/triplet $8\times8$ Nambu matrix \gf, which is the so-called {\em
  dynamical mean field theory} (DMFT)
\cite{MuellerHartmann89,Pruschke95,Georges96}  analog
of the Eliashberg equation. However, no Migdal theorem is required
since the full local irreducible pairing vertex is used.  This
non-perturbative set of equations lead to  quasi-particle and
anomalous \gf s which are gapped at $\w=0$, as shown in Sec.\  \ref{sec:tt-gf}.
The gap function is
non-analytic in frequency, and, therefore, Heid's theorem \cite{Heid95}
is not applicable. This superconducting phase belongs to a
minimum of the free energy.

\section{Theory}
\label{sec:theory}
\subsection{Model}

Uranium based \HF\ materials exhibit a lot of interesting and unusual physical
properties: coexistence of magnetism with superconductivity and \NFL\
behaviour  in the paramagnetic phase. The Uranium 5f-shell 
states in \HF\ compounds  can be
$5f^1$ (U$^{5+}$), $5f^2$ (U$^{4+}$) and $5f^3$
(U$^{3+}$). Experimentally, however, U$^{5+}$ is energetically unlikely.
The Hund's rules ground state multiplets are given by $J=5/2$, $J=4$
and $J=9/2$ respectively and  split in a cubic symmetry
\cite{LeaLeaskWolf62} according to the
\irrep s of the point group. If we
consider only valence fluctuations between the lowest \irrep s of $5f^2
\leftrightarrow 5f^3$ or $5f^2 \leftrightarrow 5f^1$, the Uranium
$5f$-shell is modelled by a magnetic doublet $\Gamma_6(\Gamma_7)$
above a non-Kramers $\Gamma_3$ doublet ground state which has a quadrupolar,
and hence orbital character. Only
conduction electrons with $\Gamma_8$ symmetry, labelled by
a spin $\sigma$ and a locally defined orbital index $\alpha$,  couple to these ionic
states according to a simple group-theoretical selection
rule  \cite{Cox87,CoxZawa98} . The same arguments also apply to
$Pr^{3+}$ ($J=4$) and $Pr^{4+}$ ($J=5/2$) in a cubic crystalline
environment which is found in the skutterudite materials
PrFe$_{4}$P$_{12}$, which shows anitferro-quadrupolar order around 2K,
and PrOs$_{4}$Sb$_{12}$ where a superconducting transition was found
at $2K$ \cite{BrianEric2002}. If we restrict ourselves to local
hybridization and coupling to two generate band - an assumption which
we will justify within the DMFT
the lattice model 
on a cubic lattice
\begin{eqnarray}
\label{eq:tca-97}
  \label{eq:hamil-tc-pam}
  H &=& \sum_{ \k\sigma\alpha} \e_{\k\sigma\alpha} 
c^\dagger_{\k\alpha\sigma}c_{\k\alpha\sigma}
+ \sum_{\sigma\nu} E_\sigma X^{\nu}_{\Gamma_6\sigma,\Gamma_6\sigma}
+ \sum_{\alpha\nu} E_\alpha X^{\nu}_{\Gamma_3\alpha,\Gamma_3\alpha}
\nonumber\\
&&   
+ \frac{1}{\sqrt{N}}  \sum_{\nu\alpha\sigma}
\left( 
 V_{\alpha}
  c^\dagger_{\nu\alpha\sigma }  X_{\Gamma_3 -\alpha,\Gamma_6\sigma}^{\nu}
  + h.c \right)
\end{eqnarray}
has an  $SU(2)\otimes SU(2)$ symmetric with respect the spin $\sigma$
and channel $\alpha$ space. $X^\nu_{\gamma,\gamma'}$ are the usual
Hubbard projection operators \cite{Hubbard64} at lattice site $\nu$
which describe transitions from local $4(5)f$ atomic state $\gamma'$ to
the state $\gamma$.

The Hamiltonian (\ref{eq:hamil-tc-pam}) contains three different energy
scales: the band-width $D$ of the conduction bands $\e_{\as}$, the
inter-configurational energy $\e_{\as} = E_\sigma - 
E_\alpha$ and the so-called Anderson or hybridization width
 $\Delta\equiv \pi \rho_0 V^2$. $\rho_0$ is the conduction electron
 density of states at the chemical potential $\mu=0$.  If not otherwise stated, we
will use $\Delta$ as the energy-scale throughout this paper.

\subsection{Dynamical Mean Field Theory}

The fundamental assumption of the DMFT is the locality of 
single particle self-energy which allows the
mapping of a complicated lattice problem onto an effective site problem
\cite{Grewe87,Kuramoto85} which is embedded in a fictitious  bath of
conduction electrons. This bath, the so-called media, takes into account
the angular averaged but energy dependent influence of the lattice
onto this effective site. The DMFT self-consistency condition (SCC) demands
the equality of the $\k$-summed lattice  Green function with the Green
function  of the effective site. Then,
only one-particle irreducible loops, which describe the  hopping of a
single electron from and to the effective site, enters the SCC. Any
two particle irreducible loops - correlated hopping of two electrons
from and back onto the site - are factorized in a product of
one-particle loops: higher order correlations which also contribute to
a local self-energy, are neglected.  This assumption becomes exact in the
limit of infinity spatial dimensions
\cite{MetznerVollhardt89,Pruschke95,Georges96}. 

Even if we had taken into account a band dispersion
$\e_{\k\sigma\alpha\alpha'}$, non-diagonal in the orbital
index $\alpha$, the
mapping onto an effective site requires 
two locally defined, degenerate effective conduction bands which are
symmetry projected linear combinations of the Wannier state of the
conduction bands. Cox has shown that only those hybridize with the
local $f$-states on symmetry grounds \cite{Cox87}. Since the local
dynamics is governed by the correlations induced by the $f$-states, it
is consistent with the DMFT to restrict ourselves to diagonal band dispersions
$\e_{\k\sigma\alpha}$ in Eqn.\ (\ref{eq:hamil-tc-pam}).
We use the so-called Non-Crossing Approximation (NCA)
\cite{Grewe83,Kuramoto83,Bickers87} 
as a solver for the effective impurity throughout the paper since it  has
been established as standard technique for solving the
two-channel lattice model \cite{AndersJarCox97,Anders99}.

\subsubsection{Single-Particle \GF}
\label{sec:s-p-gf}

The conduction electron \gf\  $G_{\alpha\sigma}(\k,z)$ for the model
(\ref{eq:hamil-tc-pam}) obeys the exact equation of motion:
\begin{equation}
  \label{eq:conduction-electrons}
  G_{\alpha\sigma}(\k,z) = \frac{1}{z-\e_{\kas}} + 
\frac{1}{z-\e_{\kas}} |V_{\alpha}|^2
F_{\alpha\sigma}(\k,z)\frac{1}{z-\e_{\kas}}  
\komma
\end{equation}
which connects it to the $f$-\gf
\begin{equation}
  \label{eq:f-gf}
  F_{\alpha\sigma}(\k,z) = 
\frac{1}{N} \sum_{i} e^{i\k \vec{R}_i}
\corrz{ X_{-\alpha,\sigma}^{i}}{X_{\sigma,-\alpha}^{i}}
\punkt
\end{equation}
Since $F_{\alpha\sigma}(\k,z)$ can always be written as
\begin{equation}
  \label{eq:f-anatz-gf}
  F_{\alpha\sigma}(\k,z) = 
\left[\bar F^{-1}_{\as}(\k,z)    -\frac{|V_{\alpha}|^2}{z-\e_{\kas}}
\right]^{-1} 
\komma
\end{equation}
where in DMFT $\bar F^{-1}_{\as}(\k,z) = \bar
F^{-1}_{\as}(z) = z -\e_{\as}  
-\Sigma^f_{\as}(z)$, the DMFT solution
for the single-particle \gf\ of 
two-channel  periodic Anderson model is determined by the solution of
\begin{eqnarray}
  \label{eq:tca-scc}
  F_{loc,\as}(z) & = &  \frac{1}{z-\e_{\as} -\Delta_{\as}(z)
    -\Sigma^f_{\as}(z)}
\\
  &=& \frac{1}{N} \sum_{\k} 
\left[\bar F^{-1}_{\as}(z)    -\frac{|V_{\alpha}|^2}{z-\e_{\kas}}
\right]^{-1} 
\nonumber
\punkt
\end{eqnarray}
The local \se\ $\Sigma^f_{\as}(z)$ is calculated by a
SIAM with an effective conduction-electron density of states 
$\tilde\rho(\w) = \Im m \Delta(\w-i\delta)/|\pi V_{\alpha}^2|$. Therefore,
the \se\ of the conduction electrons  $\Sigma_{\alpha\sigma}^{c}(\k,z)
= z -\e_{\kas} - G_{\alpha\sigma}^{-1}(\k,z)$ is given by
\begin{equation}
  \label{eq:self-energy-gc}
  \Sigma_{\alpha\sigma}^{(c)}(z) = 
\frac{ T_{\as}(z) }{1 + \tilde G  T_{\as}(z)} =
  \frac{|V_{\alpha}|^2}{ z -\e_{\as} -\Sigma^f_{\as}(z)} 
\end{equation}
as a consequence of the equation of motion
(\ref{eq:conduction-electrons}) where $T_{\as}(z) = |V_{\alpha}|^2 \bar
F_{\as} (z)$.
Eqns.~(\ref{eq:conduction-electrons}-\ref{eq:self-energy-gc}) are  
valid for any number of channels. 

In contrast to the single channel Kondo effect, the ground
state cannot be mapped onto a renormalized Fermi liquid in the two
channel case. Here, the effective
coupling constant flows to an intermediate value. The
corresponding fixed point has \NFL\ properties and is unstable against a
magnetic and a channel field \cite{PangCox91}.

The single particle and transport properties of this model
\cite{AndersJarCox97}  show
\NFL-behaviour  within the DMFT approximation which can be seen
immediately by the  analytic form of the conduction electron
self-energy (\ref{eq:self-energy-gc}): independent on the method which
is used to solve the effective impurity, the local $f$-\se\
$\Sigma^f_{\as}(z)$ and, therefore, also  self-energy of the conduction
electrons, must have a finite imaginary part even for
$T\rightarrow 0$ in the paramagnetic phase  since the local $T$-matrix is
pinned to  half of the  unitary limit
\cite{CoxZawa98,PangCox91,EmeryKiv92}. The scattering originates in the
two-channel 
Kondo effect in the impurity model \cite{Cox87}: the local quadrupolar moments of the $\Gamma_3$
states are dynamically screened by the conduction electrons
orbital moments, 
if $E_\alpha< E_\sigma$. In the case of $E_\alpha >E_\sigma$, orbital and spin
degrees of freedom are interchanged, and the local magnetic moments are
screened instead. 
The predicted excitation gap at low temperatures in the optical conductivity
\cite{AndersJarCox97} has been found in UBe$_{13}$
\cite{BommeliDegiorgi97}.

This result of a finite resistivity in the paramagnetic
phase, originating in  magnetic scattering at the $f$-sites due to the
absents of singlet  ground state, indicates the instability of
this  phase towards those phase transitions which will
remove the residual  
entropy of the lattice. We showed in a previous publication
\cite{Anders99}, that the model has a rich phase diagram of
spin-density, quadrupolar-density, and anti-ferromagnetic and
anti-quadrupolar order for larger $|\e_{\as}|>\Delta$. For
the intermediate valent regime ($|\e_{\as}|<\Delta$) ferromagnetism is
found for small band fillings. We suspect, that superconductivity
dominates the intermediate valent regime.

\begin{figure}[t]

\includegraphics[width=80mm]{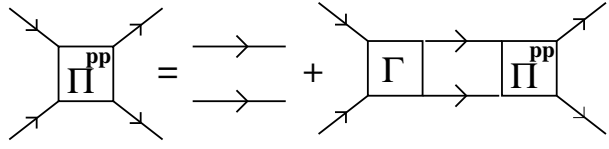}
    \caption{Bethe-Salpeter equation for the two-particle \gf\  $\Pi^{pp}$,
      $\Gamma$ denotes the irreducible vertex.}
    \label{fig:two-particle-prop}
\end{figure}

\subsubsection{Particle-Particle Green Function}
\label{sec:pp-gh}

The main focus of this investigation is devoted to the study  of
superconducting transitions. One indicator of a phase transition is a
divergence 
of the static order parameter \sus. Therefore, we are
interested in Fourier components of space-time \suss\ 
of the type 
\begin{equation}
  \label{eq:sus-def}
  \chi_{O_i,O_j}(t) = -<T(O_i(t) \; O^\dagger_j)>
\; \komma
\end{equation}
where $O_i$ will be a symmetry adapted Cooper pair operator
$c_{i\as}c_{i\apsp}$. The  lattice \suss\ are
obtained as
\begin{eqnarray}
  \label{eq:sus-two-particle-prop}
  \chi_{O}(\q,\nu) &=& \frac{1}{N^2\beta^2}\sum_{\iwn,\k,\iwm\kp}
F_O(\iwn,\k) F^*_O(\iwm,\kp) \\
&& \times
\Pi(\iwn,\k;\iwm,\kp)(\q,\nu)
  \nonumber
\komma
\end{eqnarray}
where $F_O(\iwn,\k)$ is the Fourier component of the structure factor of
the operator $O_i(t)$ and $\Pi(\iwn,\k;\iwm,\kp)(\q,\nu)$ denotes the
two-particle \gf. The summation over spin and channel indices has
been omitted in the following section for clarity. It will be
introduced and discussed in Sec.\ \ref{sec:composite-pp}.

The \twop\ \gf\  $\Pi$ obeys a Bethe-Salpeter equation,
diagrammatically depicted in Fig.~\ref{fig:two-particle-prop}.  The
lines are fully renormalized one-particle \gf s. 
A scaling analysis  has proven  that the total irreducible vertex 
has to be  purely local in the limit of infinite spatial dimensions
 \cite{BrandMilsch89,ZlaticHorvatic90}. Total
irreducibility means that each diagrammatic part  remains
connected when any two electron lines are cut.
Although, we use the textbook definition of \twop\
irreducibility with respect to Feynman perturbation theory, the
structure of the equations are invariant of the underlying
perturbation theory since the Bethe-Salpeter equation is the
\twop\ equivalent to the Dyson equation.

It was  noticed early  in the development of the
DMFT \cite{MuellerHartmann89,ZlaticHorvatic90} that 
only the local part of the irreducible
two-particle vertex enters the
Bethe-Salpeter  equation for the \twop\ \gf. Moreover, the
renormalization of the \onep\ self-energy by the \twop\ vertex
is already taken into 
account in DMFT \cite{Pruschke95,Georges96}.  This statement, however,
is only true inside  the homogeneous (paramagnetic) phase. The \sus\
becomes divergent near the phase transition temperature $T_c$ where at
least one irreducible vertex diverges and  becomes 
$\k$-dependent. The  scaling argument  fails in this case. Therefore, the 
self-energy $ \Sigma$ acquires  an additional, possible $\k$-dependent,
contribution in the symmetry broken phase. 
This has been widely used to
investigate possible magnetic and superconduction phases in various
models \cite{JarrellPangCox97,Jarrell95,ObermeierPruschke97}.

With a local irreducible pp-vertex $\Gamma$ in the  homogeneous
phase, the 
particle-particle \gf\  $\Pi$ reads
\begin{eqnarray}
  \label{eq:two-prop-interaction}
&&  \Pi(\iwn,\k;\iwm,\kp)(\q,\ivn)
 = \non
&&
\hspace{10mm}
\beta
  N\delta_{\k,\kp}\delta_{n,m}
\Pi_{free}(\k,\q,\iwn,\ivn)
\\  
&& 
\hspace{10mm}
+ \Pi_{free}(\k,\q,\iwn,\ivn)
I(\iwn,\iwm,\ivn,\q \, )
\non
&& 
\hspace{10mm}
\times
\Pi_{free}(\kk,\q,\iwm,\ivn)
\komma
%%G_{\kp}(\iwm+\ivn)G_{\kp+\q}(\iwm)
\end{eqnarray}
where
\begin{eqnarray}
\Pi_{free}(\k,\q,\iwn,\ivn)  =  G_{\k +\q}(\iwn+\ivn)G_{-\k}(\iwn)  
\end{eqnarray}
and  the reducible vertex $I(\iwn,\iwm,\ivn,\q \, )$ obeys the
Bethe-Salpeter equation
\begin{eqnarray}
  \label{eq:reduc-inter}
 && I(\iwn,\iwm,\ivn,\q \, ) = \Gamma(\iwn,\iwm;\ivn)
%%\hspace{60mm}
\\  
&&
\hspace{10mm}
+
\frac{1}{\beta}\sum_{\iw{n'}}\Gamma(\iwn,\iw{n'};\ivn) 
\non
&& 
\hspace{10mm}
\times
\chi_{0}(\iw{n'};\ivn,\q \, )  I(\iw{n'},\iwm)(\ivn,\q \, )
\nonumber 
\punkt
\end{eqnarray}
Herein, the particle-particle bubble $\chi_{0}(\iw{n};\ivn,\q \, )$
is given by 
\begin{eqnarray}
  \label{eq:pp-bubble}
  \chi_{0}(\iw{n};\ivn,\q \, ) & = & \frac{1}{N}\sum_{\k}
\Pi_{free}(\k,\q,\iwn,\ivn) 
\punkt
\end{eqnarray}
Energy $\ivn$  and  momentum $\q$ of the center of mass (COM)
are conserved.  It was shown \cite{MuellerHartmann89} that
the COM-momentum only enters through a scalar $\eta(\q \, )$
\begin{equation}
  \label{eq:define-of-x-q}
  \eta(\q \, ) = \frac{1}{d} \sum_{\alpha}^{d} \cos(q_\alpha)
\end{equation}
on a hyper-cubic lattice in $d$ dimensions. In this case, Eqn
(\ref{eq:reduc-inter}) is 
casted into a matrix equation in  Matsubara-frequency space
\begin{equation}
  \label{eq:I-matrix}
  \mat{I}(\ivn,\q \, ) = \mat{\Gamma}(\ivn) + \frac{1}{\beta}
  \mat{\Gamma}(\ivn)\mat{\chi}_0(\ivn,\q \, )
  \mat{I}(\ivn,\q \, )
\komma
\end{equation}
which is  formally solved by
\begin{equation}
  \label{eq:I-matrix-solution}
  \mat{I} (\ivn,\q \, )= 
\left[1 - \frac{1}{\beta}
  \mat{\Gamma}(\ivn)\mat{\chi}_0(\ivn,\q \, )
\right]^{-1}\mat{\Gamma}(\ivn)
\punkt
\end{equation}

The DMFT mapping of the lattice problem onto an effective impurity problem
is used in order  to calculate the irreducible vertex
$\mat{\Gamma}^{ph}(\ivn)$. 
In formal equivalence to Eqn.~(\ref{eq:I-matrix}),
the equation
\begin{equation}
  \label{eq:local-ph-prop}
  \mat{\Pi}_{loc}(\ivn) = \beta \mat{\chi}_{loc} +
  \mat{\chi}_{loc} \frac{1}{\beta} \mat{\Gamma}(\ivn)
  \mat{\Pi}_{loc}(\ivn)  
\komma
\end{equation}
depicted in Fig.~\ref{fig:two-particle-prop}, 
holds for the local two-particle \gf\  of the effective site. 
It has been shown that the
irreducible vertices $\mat{\Gamma}(\ivn)$ in
(\ref{eq:local-ph-prop}) and in (\ref{eq:I-matrix-solution}) are identical.
They may be obtained by
matrix inversion in Matsubara frequency space
\begin{equation}
  \label{eq:irr-ph-vertex}
  \frac{1}{\beta} \mat{\Gamma}(\ivn) =
  [\mat{\chi}_{loc}]^{-1} - \beta [\mat{\Pi}_{loc}(\ivn) ]^{-1}
\;\; 
\end{equation}
once the free local pp-\gf\  
\begin{equation}
\chi_{loc}(\iwn,\iwm;\ivn) = 
G(\iwn+\ivn)G(-\iwn)\delta_{\iwn,\iwm}
\end{equation}
 and the local pp-\gf\ $\mat{\Pi}_{loc}(\ivn)$ are known.

\subsubsection{Local Two-Particle \GF}
\label{sec:local-two-part-prop}

\begin{figure}[tb]
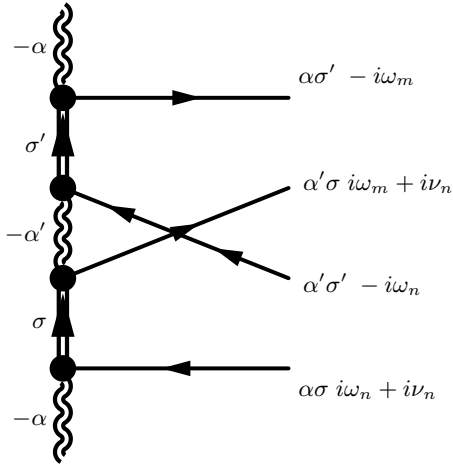

  \begin{center}

    \begin{fmfchar*}(30,60)
%%     \fmfpen{thick}
      \fmfpen{0.8thick}
      \fmfstraight
      \fmfleftn{v}{6}
      \fmfrightn{c}{6}
      \fmf{fermion}{c2,v2}
      \fmf{fermion}{c3,h1,v4}
      
      \fmflabel{$\apsp\; -\iwn$}{c3}
      \fmflabel{$\as\; \iwn+\ivn$}{c2}
      
      \fmflabel{$\asp\; -\iwm$}{c5}
      \fmflabel{$\aps\; \iwm+\ivn$}{c4}
      
      \fmf{fermion}{v3,c4}
      \fmf{fermion}{v5,c5}
      
      \fmf{dbl_wiggly,label.side=left,label=$-\alpha$}{v1,v2}
      \fmf{dbl_plain_arrow,label.side=left,label=$\sigma$}{v2,v3}
      \fmf{dbl_wiggly,label.side=left,label=$-\alpha'$}{v3,v4}
      \fmf{dbl_plain_arrow,label.side=left,label=$\sigma'$}{v4,v5}
      \fmf{dbl_wiggly,label.side=left,label=$-\alpha$}{v5,v6}
      \fmfmydot{0.1w}{v2}     
      \fmfmydot{0.1w}{v3}     
      \fmfmydot{0.1w}{v4}     
      \fmfmydot{0.1w}{v5}     
    \end{fmfchar*}

    \caption{Local particle-particle  \Gf\  for the \tc\  model
      in NCA plotted on an imaginary time axis, from bottom to
      top. The wiggly vertical line represents the $\Gamma_3$ states, the
      double straight the magnetic $\Gamma_6$ states and the horizonal
      lines the hybridizating conduction electrons.}
    \label{fig:local-pp-prop}
  \end{center}
%%\end{figure}
\end{figure}

The local two-particle $f$-\gf\  $\Pi_f(\iwn,\iwm;\ivn)$
for the periodic Anderson model has been calculated within
the NCA  in the limit of two local valence configurations, one with $N$
electrons and one with $N+1$ electrons \cite{Grewe97}. 
For the two-channel model, the diagrammatic contribution 
is shown in Fig.\ \ref{fig:local-pp-prop} and reads:
  \begin{eqnarray}
\Pi_f(\iwn,\iw{m};\ivn)_{\as;\apsp}^{\aps;\asp}  = 
\nonumber \\ 
-\myCint P_{-\alpha}(z)P_{\sigma}(z+\iwn+\ivn) 
\nonumber \\ 
%%& &
 \times 
P_{-\alpha'}(z+\iwn-\iw{m})
  P_{\sigma'}(z-\iw{m})
\label{equ-pp-f-ex}
  \end{eqnarray}
where the contour circumvents all singularities of the
ionic propagators $P_{\gamma}(z)$. $\tilde Z_{eff}$ is the partition
function of the effective site.
The indices on the bottom of $\Pi_f$ label the spin and channel
of the incoming while the indices on the top label the outgoing particle-hole
pair. Since there is one  external
line crossings when $\Pi_f^{pp}$ is inserted into the Bethe-Salpeter Eqn.\
(\ref{eq:local-ph-prop}), the overall sign must be negative.
\def\sign{\mbox{sign}}

In general, Eqn.~(\ref{equ-pp-f-ex}) has to be numerically  evaluated. 
At $T=0$, however, this integral can be solved analyticly for $|\iwn|
\ll \Delta$ for the two-channel Anderson model
if we use the  ionic propagators  \cite{MullerHartmann84},
assuming that the lattice self-consistency does not change the
low-frequency threshold behaviour.
This is justified in the
paramagnetic phase for an energy independent DOS close to $\w=0$. A rather
complicated calculation \cite{CoxZawa98} yields 
\begin{eqnarray}
\label{equ-pi-f-ex-nca}
\Pi_f(\iwn,\iwm;0) & \approx&
 \left(\frac{\pi}{4\Delta}\right)^2
\frac{\sign(\w_n)\cdot \sign(\w_m)}{|\max(\w_n,\w_m)|}
\nonumber \\ 
&& \times
F\left(\frac{\pi}{2}; \left|\frac{\min(\w_n,\w_m)}{\max(\w_n,\w_m)}
\right|\right)
\komma  
\end{eqnarray}
where
$F(\frac{\pi}{2},x< 1)$ is the elliptic integral of
the first kind. 
For low frequencies $\iwn$ and $\iwm$, $\Pi_f(\iwn,\iwm;0)$ turns out
to be independent of the low energy 
scale $T_K$. The conduction electrons
over-screen the local spin: a compensation of the spin can never be
achieved, and the conduction electrons continue to be incoherently scattered
 for $T\rightarrow 0$. The \twop\ scattering is
governed by  low lying excitations 
independent of the cross-over energy scale $T_K$, and the temperature
$T$ appears to be the only 
relevant energy scale. This approximation is valid for $|\w|, T <T_K$.

Since the large Coulomb repulsion prevents the ionic shell from
accepting or donating two electrons at the same time,
a local pair correlation at site $i$  can only be generated  sequentially.
The first electron has to leave the $f$-shell  before another
conduction electron can enter it,  see Fig.~\ref{fig:local-pp-prop}.
Therefore, the particle-particle $f$-\gf\  tends to have nodes as a
function of time. $\Pi_f(\iwn,\iwn;0)$ becomes an odd function
of both of its frequency arguments $\iwn$ and $\iwm$ in the
two-channel model due to the two exchangeable doublets. 
In other models with locally
restricted charge fluctuations, we
expect that odd and even  components with respect to frequency will be present.

The \pam\ contains two types of electrons: localized and itinerate
ones.  
Since 
the bands are non-interacting in the absence of
hybridization, particle-particle correlations  only occurs through
hybridization with localized  states in ionic shells. We can express
the conduction-band particle-particle \Gf\ in DMFT as
\begin{equation}
  \label{eq:eq-ph-cc-propagator}
\Pi_{c}(\iwn,\iwm;\ivn)  =
\beta \chi_{c,loc}
+ \tilde  \chi_{c,loc} {\cal T}_{loc}
\tilde  \chi_{c,loc}
\komma
\end{equation}
where energy arguments have been dropped on the r.h.s.
The local \twop\ $T$-matrix ${\cal T}_{loc}^{(ph)}$ cumulant is defined by
\begin{eqnarray}
\label{eqn:local-ph-t-mat}
{\cal T}_{loc}(\iwn,\iwm;\ivn) &=&
V^4 (\Pi_f(\iwn,\iwn;\ivn)  
\nonumber \\
&&
- \beta F(-\iwn)F(\ivn+\iwn)
\delta_{n,m})
\punkt  
\end{eqnarray}
$\chi_{c,loc}(\iwn,\iwm;\ivn) =  \delta_{n,m} G_c(-\iwn)
G_c(\iwn+\ivn)$,
and the media  particle-particle \gf\ is 
$$
\tilde  \chi_{c,loc} =  \tilde
G(-\iwn) \tilde G(\iwn+\ivn)
\punkt
$$
$\Pi_f$ factorizes in a product
of two $f$-\gf s 
in the absence of local interactions, and the \twop\ $T$-matrix cumulant
vanishes as expected. 
Using Eqn.~(\ref{eq:irr-ph-vertex}) and
(\ref{eq:eq-ph-cc-propagator}) we obtain 
\begin{eqnarray}
\label{eq:irr-vertex-dmft}
\frac{1}{\beta}\mat{\Gamma}_c 
&  = &
\mat{A}
\frac{1}{\beta} \mat{{\cal T}}_{loc}
\mat{A}
\frac{1}{
1 + 
\mat{\tilde\chi}_{c,loc}
\frac{1}{\beta} \mat{{\cal T}}_{loc}
\mat{A}
}
\\
\mat{A} & = & \mat{\tilde\chi}_{c,loc}
[\mat{\chi}_{c,loc}]^{-1}
\punkt  
\end{eqnarray}
The matrix $\mat{A}$ measures the degree of renormalization of the
band electrons. It is of
order $O(1)$ at high temperatures and becomes moderately enhanced in
the low temperature regime. Hence, the \irr\ vertex
$\mat{\Gamma}_c \approx \mat{{\cal T}}_{loc}$.
If we define a renormalized $T$-matrix 
$\mat{\tilde { \cal T}}_{loc} = \mat{A} \; \mat{{\cal T}}_{loc}
\;  \mat{A}$, the \irr\  vertex assumes the same structure 
as the \onep\ self-energy $ \Sigma_{\alpha\sigma}^{(c)}(z)$
(\ref{eq:self-energy-gc}) 
 \begin{equation}
   \label{eq:irr-vertex-t-matrix}
   \mat{\Gamma}_c = \mat{\tilde {\cal T}}_{loc} 
\left[\mat{1} + \mat{\chi}_{c,loc}\mat{\tilde {\cal
      T}}_{loc}  \right]^{-1}
\punkt
 \end{equation}

\section{Superconductivity}
\label{sec:superconductivity}

\subsection{Classification of the Cooper-Pair Operators}
\def\nn{\\ }

It is assumed that
only the electrons in the conduction electron sub-system are directly
involved in the formation of Cooper pairs in the condensed phase. The quantum numbers $\gamma$ and $\gamma'$  label  their spin
and orbital degrees of freedom. If a Cooper  pair has an internal 
structure modelled by a structure factor $S(2\vec{r})$, its pair operator reads
\begin{equation}
  \label{eq:cooper-pair}
  P_{\gamma,\gamma'}(\vec{R}) = \sum_{\vec{r}} S(2\vec{r})
  c_{\gamma}(\vec{R}-\vec{r})
c_{\gamma'}(\vec{R}+\vec{r})
\punkt
\end{equation}
$\vec{r} = (\vec{r}_1 -\vec{r}_2)/2$ determines the location of one
electron relative to the center of mass $\vec{R} = (\vec{r}_1 +\vec{r}_2)/2$.
The transformed pair operator $P_{\gamma,\gamma'}(\vec{q})$
\begin{equation}
  \label{eq:cooper-pair-com-q}
  P_{\gamma,\gamma'}(\vec{q}) = \sum_{\vec{k}} S(\vec{k})
  c_{\gamma}(\vec{k}-\vec{q}/2)
c_{\gamma'}(-\vec{k}-\vec{q}/2)
\end{equation}
describes a pair  in reciprocal space with a center of mass
momentum $\vec{q}$.

\begin{table}[t]
  \begin{center}
    \begin{tabular}{|r|l|l|c|c|}
      \hline
    Sector &  Spin & Channel & Frequency & \# Orderp. \\ %%& $F(z) = s F(-z)$ \nn
      \hline 
      \hline
    I (SsCs) &  singlet & singlet & odd  & 1  \nn
   II (SsCt) &   singlet &   triplet & even  & 3   \nn
   III (StCc) & triplet &   singlet  & even  & 3   \nn
   IV (StCt)&  triplet  & triplet  & odd & 9   \\
      \hline
    \end{tabular}
    \caption{Classification of the 16 possible order parameters for
      local Cooper pairs.  Pauli's principle determines the frequency
      behaviour of the symmetrized anomalous \gf\ $F(z)$. Odd
      frequency implies $(F(z) = -F(-z)$, even frequency $F(z) = F(-z)$. Only Cooper
      pairs  with $S(\k)$ transforming according to the \irrep\
      $\Gamma_1$ are taken into account. Sector I and IV show odd frequency behaviour, while
      in sector II and III conventional even frequency pairing is found.}
    \label{tab:sym-order-parameter}
  \end{center}
%\end{table}
\end{table}

Since in DMFT  the irreducible particle-particle (pp) vertex
$\Gamma$ is also a $\q$-independent quantity, compare
Eqns.~(\ref{eq:two-prop-interaction}) to (\ref{eq:reduc-inter}),
the pair \sus\
\begin{equation}
  \label{eq:pair-sus}
\chi_{p,\gamma,\gamma',\gamma'',\gamma'''}(\q,\tau) = \expect{T (P_{\gamma,\gamma'}(\vec{q})(\tau)
  P^\dagger_{\gamma'',\gamma'''}(\vec{q}) ) }
\komma
\end{equation}
is only influenced by the local interaction if the structure factor
$S(\k)$ transforms with $\Gamma_1$, the trivial representation of the
point group. Hence, solely superconducting instabilities  with a conventional
order parameter which also transforms according to $\Gamma_1$  can be
obtained within DMFT through 
(\ref{eq:pair-sus}).
 This includes, however, Cooper pair form-factors
such as $S(\k) \propto \e_{\k}$, sometimes called {\em extended
  s-wave}, which will play the leading role in the extreme mixed valent
regime where $\e_{\as}=0$.
Therefore, we will ignore unconventional spatial form factor
symmetries which, of course, cannot be ruled out in the model.

The corresponding one-particle anomalous \gf\
$F_{\gamma,\gamma'}(\k,\q,z)$ is given by the Fourier transform of
\begin{equation}
  \label{eq:anomalous-gf}
F_{\gamma,\gamma'}(\k,\q,\tau) = 
-\langle  T
%\left
(
c_{\gamma,\vec{k}-\vec{q}/2}(\tau)  c_{\gamma',-\vec{k}-\vec{q}/2}
%\right
) \rangle
\komma
\end{equation}
where $\vec{q}$ denotes the momentum of the Cooper pair.

Since $\gamma$ contains the spin and orbital degrees
$\sigma\alpha$, the possible symmetry properties of the pair \sus\ can
be assigned to the
four sectors listed in table \ref{tab:sym-order-parameter}. These
are associated with certain subspaces of the product space of spin and
channel degrees of freedoms. Each of them are spanned by a product of
singlet and triplet states as indicated.
Since the total symmetry under exchange of the electrons has to be odd
due to Pauli's principle, the frequency dependence of the
symmetrized anomalous \gf\ $F(\vec{q},\iwn)$ 
\begin{equation}
  \label{eq:symmetrice-gf}
  F(\vec{q},\iwn) = \sum_{\gamma \gamma'\k} \lambda_{\gamma\gamma'} S(\k)
F_{\gamma,\gamma'}(\k,\q,\iwn)
\end{equation}
has to be odd in sector I and IV; hereby, $ \lambda_{\gamma\gamma'}$
denotes the symmetry form-factor of the appropriate sector.

The most general pair operator in $\k$-space in sector I
(spin singlet/channel singlet)  is given by
\begin{equation}
  \label{eq:pair-ss}
  P^{ss}(\k,\kk)  = S(\frac{\k+\kk}{2})
 \sum_{\alpha\sigma}\alpha \sigma c_{\k\alpha\sigma}  c_{-\kk-\alpha-\sigma}
\komma
\end{equation}
from which a symmetrized pair operator of the form
(\ref{eq:cooper-pair-com-q}) can be obtained  by
\begin{equation}
  P^{ss}(\vec{q}) = \sum_{\k} P^{ss}(\k+\vec{q}/2,\k-\vec{q}/2)
\punkt
\end{equation}

When we introduce the transposed bi-spinor
operator $\vec{\psi}^T(\k) =  \left( c_{\k+\uparrow},c_{\k+\downarrow},
c_{\k-\uparrow},c_{\k-\downarrow}\right)$, the general pair operator in sector
II (spin singlet/channel triplet) can be written as
\begin{equation}
  \label{eq:pair-s-t}
  \vec{P}^{s,t}(\k,\kk)  = 
  S(\frac{\k+\kk}{2}) \vec{\psi}^T(\k)i\mat{\sigma}_y\;
 i\mat{\tau}_{y}\mat{\vec{\tau}} \vec{\psi}(-\kk) 
\end{equation}
and  as
\begin{equation}
  \label{eq:pair-t-s}
  \vec{P}^{t,s}(\k,\kk)  = 
  S(\frac{\k+\kk}{2}) \vec{\psi}^T(\k)i\mat{\tau}_y\;
 i\mat{\sigma}_{y}\mat{\vec{\sigma}} \vec{\psi}(-\kk) 
\end{equation}
in sector III (spin triplet/channel singlet),
where $\mat{\vec{\sigma}}$ acts in the spin sector and
$\mat{\vec{\tau}}$ in the channel sector. The nine pair operators of
sector IV ( spin and channel triplet)  are given by the matrix with
the elements
\begin{equation}
  \label{eq:pair-t-t}
  P^{t,t}_{i,j}(\k,\kk)  = 
  S(\frac{\k+\kk}{2}) \vec{\psi}^T(\k)i\mat{\sigma}_{y} \mat{\sigma}_i
 \;i\mat{\tau}_y\mat{\tau}_j \vec{\psi}(-\kk) 
\punkt
\end{equation}

\subsection{Introduction to Odd-Frequency Pairing}

In section \ref{sec:local-two-part-prop}, it was shown that the
irreducible vertex $\Gamma$ is proportional to  the local
$f$-particle-particle propagator $\Pi_f$ in 
leading order, and that $\Pi_f$ is odd in the incoming and outgoing
frequencies at low temperatures, Eqn (\ref{equ-pi-f-ex-nca}). Hence,
we expect to find a tendency 
towards so-called odd-frequency superconductivity in model
(\ref{eq:hamil-tc-pam}).

Berezinskii has pointed out in the context of \hel\ that 
$s$-wave/spin triplet or $p$-wave/spin singlet pairing, both having 
a temporal node, are also compatible with Pauli's principle
\cite{Berezinskii74}: it only requires a total parity of -1. 
The anomalous \gf\ has  to be an odd function, with respect to
frequency, in the above mentioned cases.
However, it quickly became apparent  that \hel\ orders in $p$-wave
(L=1), spin triplet (S=1) pairs which show even frequency behaviour.
Berezinskii's  idea was forgotten, but was later revitalized by
Balatsky \etal\ \cite{BalatskyAbrahams92} in the context of
superconducting phases of the Hubbard-model.  Similar to magnetic
phases, a superconduction ordered phase can have an order parameter
whose modulation in real space is described by a finite wave vector
$\vec{q}$. This wave vector corresponds to a finite center of mass
momentum $\vec{q}$ of the Cooper pairs. Coleman \etal\
\cite{ColemanMirandaTsvelik94} investigated the possibility of an
odd-frequency solution for a Kondo lattice model and found
a stable solution for such a staggered order parameter within a mean-field
theory. This solution, however, breaks spin-rotational invariance. The
order parameter consists of a local 
spin contribution bound to a conduction electron and transforms as an
$S=1/2$ object.
Abrahams \etal\ \cite{Abrahams95}  argued that the order parameter of
an odd-frequency superconductor
has to be a composite object, i.~e.\ a Cooper pair  bound to a localized
spin, in order to produce a positive Meissner effect. Therefore, time
reversal symmetry is not broken by odd-frequency pairing since both
components of the order parameter, the
Cooper pair and the spin operator, have negative parity with respect
to the time-reversal operator.
Heid, however, showed  that odd-frequency
superconductors are  thermodynamically instable against $\q$-modulation
\cite{Heid95}.
He assumed that the normal-phase was described by  Fermi-liquid theory
and that the order parameter did not have singularities.  Emery and
Kivelson \cite{EmeryKiv92}  found an enhancement of a composite pair
operator \sus\  using the solution of the two-channel Kondo-model at
the Toulouse point. This could  indicate a tendency towards an
odd-frequency phase in a corresponding lattice model.

%%Even now, 
Since the mechanism of superconductivity in strongly  correlated
electron systems is still subject to an ongoing  debate, it is of
great importance to explore  this  unusual concept in particular
since, as will be seen  below, it is favoured by orbital
degrees of freedom. Whether odd-frequency pairing is actually realized
in nature, is not yet clear.

\subsection{Composite Order Parameter and its Physical Interpretation}
\label{sec:composite-orderparameter}

Pauli's principle demands that superconductivity  in the
spin singlet, channel singlet (SsCs) sector  
or in the  spin triplet, channel triplet (StCt)  with a $\Gamma_1$
form-factor must have an odd anomalous \gf\ as indicated in table
\ref{tab:sym-order-parameter}. 
Assuming pairing with $\vec{q}=0$, the local pair expectation value 
\begin{equation}
  \label{eq:local-pair-expect-ss}
  \expect{P_{ss}(\vec{q}=0)} = \frac{1}{N}\sum_{\k} \sum_{\alpha\sigma}\alpha
  \sigma S(\k)
\expect{c_{\k\alpha\sigma}  c_{-\k-\alpha-\sigma}} = 0
\end{equation}
vanishes with  the anti-commutator $\{c_{\k\alpha\sigma},
c_{-\k-\alpha-\sigma}\} = 0$, which reflects Pauli's principle. Hence,
this expectation value cannot serve as  the superconducting
order parameter of  section I and IV. This is a peculiarity of
 \oddf\ superconductivity.

Abrahams \etal\ 
noticed that the time derivative
$D_{\gamma\gamma'}=\expect{\frac{d}{d\tau}c_\gamma(\tau) 
  c_{\gamma'}}$  is non-vanishing in the odd frequency superconducting phase
and may possibly be furnished as an order parameter \cite{Abrahams95}. 
Since the time derivative operator does not change the symmetry of
the spin and orbital part, the order parameter
can still be classified according to the symmetry of the conduction
band Cooper pairs listed in table \ref{tab:sym-order-parameter}.
This new kind of  order parameter is formed by linear combinations of
$D_{\gamma\gamma'}$, which will be called $O$ below.
The derivative $\frac{d}{d\tau}c_\gamma(\tau)$ is equivalent to the
commutator $[H,c_\gamma(\tau) ]$. In the case of the \tcpam,
$\frac{d}{d\tau}c_\gamma(\tau) c_{\gamma'}$ consists of a product of
two operators: a conduction electron pair operator and a local magnetic 
or channel spin operator  both generated by the hybridization  term
between the band electrons and the local $f$-states.
Thus, one arrives at a composite order parameter, which correlates
local and itinerant degrees of freedom: formation of band Cooper
pairs is stimulated in the presence of a local pair resonating between
definite spin and channel states with proper symmetry.

Since there is only one order parameter component $O_{ss}$  in sector
I, it must be a global singlet. The only possibility to generate a
singlet from a vector operator such as  the local $f$-spin $\vec S$ is
to form a 
scalar product
involving another vector operator in the spin space. It has to be the
conduction  electron pair operator $\vec{P}^{t,s}$ given by
Eqn.~(\ref{eq:pair-t-s}) which is a scalar
in orbital and a vector in spin space:
\begin{equation}
  \label{eq:order-ss}
  O_{ss} = \mbox{sign}(\Delta E) \left(
\expect{\vec{S}\; \vec{P}^{t,s}} 
- \expect{\vec{\tau}\;  \vec{P}^{s,t}}
\right)
\punkt
\end{equation}
The second term in this expression, the first term's symmetrical
counterpart, combines a local 
$f$-shell orbital spin $\vec \tau$ and a pair  operator
$\vec{P}^{s,t}$ which is a vector in orbital space and
a scalar in spin space to form a global singlet. Both
terms are expected to contribute since the \hamil\
(\ref{eq:tca-97}) models local fluctuations between  spin and orbital
doublets at each lattice site. In this way, the ``pair-wave function'' in
sector I  realizes correlations between the dynamics of the local spin
and a conduction electron pair, which are compliant with Pauli's principle.

The dimensionless order parameter $O_{ss}$ is invariant under particle-hole
transformation and simultaneous exchange of the two local doublets.
This is also a fundamental symmetry of the \hamil. The exchange of
spin and channel spin is compensated by the sign change of $\Delta E=
E_\sigma-E_{\alpha} = \eas$  being  independent of spin and channel  
in the absence of a magnetic field or lattice distortion.
Hence, the superconducting order parameter is  not dependent on
whether the magnetic or the quadrupolar doublet has the lower energy.
Moreover, the composite order parameter meets the objective that the
superconducting state must bind local magnetic or orbital degrees of
freedom in order to quench the residual entropy present in the \tc\
model. 
 
The pair operators which enter  $O_{ss}$ in Eqn.~(\ref{eq:order-ss})
transform according to the \irrep s of sectors II and III. This
reflects the fact that Pauli's principle allows that their
expectation value may stay finite. Of course, the symmetry of the composite
order parameter is determined by the scalar product as a whole and not by the
symmetry of the individual operators; therefore, the
order parameter transforms according to sector I.
A simple mean field  decoupling of $O_{ss}$, as given by
\begin{equation}
  \label{eq:o-ss-mean-field}
  O_{ss} = \mbox{sign}(\Delta E) \left(
\expect{\vec{S}} \expect{\vec{P}^{t,s}} 
- \expect{\vec{\tau} }\expect{\vec{P}^{s,t}}
\right) \komma
\end{equation}
would  essentially treat the local  and itinerant correlations  as
independent quantities and would have to be interpreted as a mixture of
phases of sectors  II and III rather then superconductivity in sector
I. Correlations between the conduction band 
pairs and the local spin and orbital degrees of freedom are essential
for generating the unusual phase in sector I. Nevertheless, this
decoupling might still be a way to understand the experimentally
observed coexistence of magnetism and superconductivity in certain
Uranium \HF\ compounds. An
antiferromagnetic spin order, in conjunction with 
superconductivity with a zero center of mass momentum of the Cooper pair, would translate into a
modulated order parameter $O_{ss}$,  with wave vector $\q$ given by
the magnetic modulation vector. Alternatively, one could speculate
about a finite momentum Cooper-pair \cite{MartisCox2000} bound to a
modulated spin or 
orbital ordered state in such a way that the total momentum of the
order-parameter is zero.

With the  definitions
\begin{eqnarray}
D_{\sigma\sigma'}^{\alpha\alpha'}  & = & 
\frac{1}{\beta}\sum_{\iwn} e^{\iwn\delta} 
\iwn
\onepart{c_{\nu\alpha\sigma}}{c_{\nu\alpha'\sigma'}}{\iwn}
  \label{eq:isotropic-wave}
\hspace{10mm}
\\
  \label{eq:extended-s-wave}
   T_{\sigma\sigma'}^{\alpha\alpha'} 
& = &
 \frac{1}{N\beta}\sum_{\iwn,\k} e^{\iwn\delta} \iwn
\e_{\k\as}\onepart{c_{\k\as}}{c_{-\k\alpha' \sigma'}}{\iwn}
\komma
\end{eqnarray}
it is straightforward to derive  exact relations for the symmetrized
order parameter for the singlet/singlet channel in the form
\begin{eqnarray}
 O_{ss} &= & 
\frac{|\Delta E|}{V^2}
\underbrace{ \sum_{\as} \as D_{\sigma-\sigma}^{\alpha-\alpha}}_{= D_{ss}}
+\frac{ \mbox{sign}(\Delta E)}{V^2} 
\underbrace{\sum_{\as} \as T_{\sigma-\sigma}^{\alpha-\alpha}}_{=T_{ss}}
\label{equ-ss-order-parameter}
\end{eqnarray}
using  equations of motion  as sketched in appendix
\ref{sec:appendix-order-parameter}. The order parameter has two
contributions: the term $D_{ss} $ describes an isotropic local Cooper
pair, i.~e.\ $S(\k)=1$, while $T_{ss}$ is formed by nearest neighbour
pairs, i.~e.\ $S(\k)=\e_{\k}/t$,  sometimes called  {\em
  extended s-wave pairing}.  
The $\k$-summation on the r.h.s.\ of (\ref{eq:extended-s-wave}) leaves
$T_{\sigma\sigma'}^{\alpha\alpha'}$ finite  since
$\onepart{c_{\k\as}}{c_{-\k\alpha' \sigma'}}{\iwn} \propto S(\k)$ for
the case of extended s-wave  pairing.

The composite  order parameter $O_{ss}$ contains  no natural
distinction between  
isotropic and anisotropic pairs with a form-factor $\e_{\k}/t$. It is
interesting to note that for the \IV\ regime 
$ |E_\sigma-E_{\alpha}|\rightarrow 0$, the nearest neighbour Cooper
pairs with {\em extended s-wave symmetry} dominate since the first
term on the r.h.s.\ of (\ref{equ-ss-order-parameter}) vanishes for
$|\Delta E| \rightarrow 0$. This can be
interpreted as condensation of delocalized pairs.
There will  not only be  nodes  in frequency in this regime but also nodes
in $\k$-space. 
%%The specific heat will have an
%%algebraic temperature behaviour and not be
%%exponentially suppressed for $T\rightarrow 0$. 
%%

An  exact equation similar to
(\ref{equ-ss-order-parameter})  can be derived for the tensor
order parameter in the triplet-triplet
sector
\begin{eqnarray}
 O_{ij} &= & 
\mbox{sign}(\Delta E)\left( \expect{s_i P^{s,t}_{j}}
- \expect{\tau_j P^{t,s}_{i}}\right)
\\
 &= & 
\frac{|\Delta E|}{V^2} D_{i,j}
+\frac{ \mbox{sign}(\Delta E)}{V^2} 
T_{ij}
\komma
\label{equ-tt-order-parameter}
\end{eqnarray}
where $i,j=x,y,z$, $i(j)$ indexes the spin(channel) space,
$P^{s,t}_{j}$ or $P^{t,s}_{i}$ is a component of the singlet/triplet
or triplet/singlet pairs (\ref{eq:pair-s-t}, \ref{eq:pair-t-s})  and the symmetrized 
isotropic ($D_{ij}$) and nearest neighbour pair ($T_{ij}$)
expectation values are obtained   
by the symmetrization rules (\ref{eq:pair-t-t}) in combination with
(\ref{eq:isotropic-wave}) and (\ref{eq:extended-s-wave}):
 \begin{equation}
 D_{ij} (T_{ij}) = \sum_{\sigma\sigma´}^{\alpha\alpha'}
D_{\sigma\sigma'}^{\alpha\alpha'}  (T_{\sigma\sigma'}^{\alpha\alpha'})
[i\mat{\sigma}_{y} \,
\mat{\sigma}_i]_{\sigma\sigma'}
[i\mat{\tau}_y \, \mat{\tau}_j ]_{\alpha\alpha'}
\punkt   
 \end{equation}

\subsection{Composite Order Parameter and Pair-\Sus}
\label{sec:composite-pp}

As explained above, the order parameter  is formed by a composite
operator combining a conduction band 
pair operator in singlet/triplet channel with a local spin or channel
operator in the two odd-frequency sectors I
and II. It measures the correlation between a local
spin and a conduction electron pair. To verify the existence of such an
unconventional phase in the model, it is necessary to either show that
there is a self-consistent solution of the order parameter equation in the
symmetry broken phase or search for divergencies of the
order parameter \sus\ $\chi_{O}(\q \, ) =
\onepart{O}{O^\dagger}{\ivn=0,\q}$.
It is cumbersome, however, to obtain a
tractable expression for $\chi_{O}(\q \, )$ since it involves a
three-particle \gf\ due to the composite nature of $O$. On the other
hand, Eqns.\
(\ref{equ-ss-order-parameter}) and (\ref{equ-tt-order-parameter}) 
connect the expectation value of the order parameter to the derivative
of the symmetrized pair operator $\frac{d}{d\tau}c^\dagger_\gamma(\tau) 
c^\dagger_{\gamma'}$ which takes the form
$P^{odd}_{\gamma\gamma'}(\iwn)= \iwn c_\gamma(\iwn)
c_{\gamma'}(-\iwn)$ in the Matsubara frequency space. Therefore, it is
equivalent to investigate the pair 
\suss\
  \begin{equation}
    \label{eq:odd-pair-sus}
\chi_{P}(\q) =
  \frac{1}{\beta^2}\sum_{\iwn,\iwm}\onepart{P^{odd}(\iwn)}{P^{\dagger\,
      odd}(\iwm)}{\ivn=0,\q}
\punkt
  \end{equation}
They are obtained from  two-particle \gf s whose four
external lines are connected to a frequency dependent and, in the case
of {\em ``extended s-wave''} $\k$-dependent vertex, generated
by $P^{odd}_{\gamma\gamma'}(\iwn)$ and
$P^{\dagger\,odd}_{\gamma\gamma'}(\iwm)$.

Up to now, all statements in section \ref{sec:superconductivity} are exact and independent of any 
approximation. In the following, however, we will calculate the
two-particle \gf s only within the DMFT. 
Knowing the irreducible \pp\ vertex $\mat{\Gamma}$ for the
appropriate symmetry, and only considering isotropic pairs,
 the pair \sus\ reads in general
\begin{eqnarray}
\label{equ-pair-susceptibility} 
\chi_{P}(\q,\ivn) &=& 
 \frac{1}{\beta}\sum_{\iwn\iwm} \iwn (-\iwm)
\left[\mat{\chi}(\q,\ivn) \right. 
\\
&& \times \left .
 \frac{1}{\mat{1} - \frac{1}{\beta}
\mat{\Gamma}(\ivn)\mat{\chi}(\q,\ivn)}
\right]_{\iwn\iwm}
\punkt   
\nonumber
\end{eqnarray}
Since $\mat{\Gamma}$ is local within the DMFT, only gap functions
transforming according to $\Gamma_1$ symmetry can be investigated. 
All other symmetries would require  a microscopic solution of the
model beyond the DMFT or the
restriction to 
simple BCS mean-field theories. Especially latter approaches almost
always confirms the ansatz and the assumptions made at the beginning of
the calculations, and, therefore, it is not clear to us whether real
physical insight is gained.

The phase transition temperature $T_c$ and the frequency dependence of
the anomalous self-energy $\Phi(\iwn)$ is obtained by solving the
eigenvalue equation
\begin{equation}
  \label{eq:base-m}
  \mat{M} \ket{\lambda,M} = \lambda \ket{\lambda,M}
\komma
\end{equation}
where the symmetric matrix
$\mat{M}$  is defined as
\begin{equation}
  \label{eq:def-mat-M}
  \mat{M} = \sqrt{\mat{\chi}(\q,0)}
\frac{1}{\beta}\mat{\Gamma}(0)
\sqrt{\mat{\chi}(\q,0)}
\punkt
\end{equation}
The divergent contribution to the pair \sus\ at $\ivn=0$, i.~e.\
\begin{equation}
  \label{eq:pair-sus-lambda}
   \chi_{P}(\q \, ) = \underbrace{\frac{1}{\beta} c_{\lambda_m}^2 \frac{1}{1-\lambda_m}}_{ P(\q)}
+ \mbox{regular terms}
\end{equation}
stems from the largest eigenvalue $\lambda_m\rightarrow 1$.
Hereby, $ c_{\lambda_m}$ is given by $\bra{\lambda}
\sqrt{\mat{\chi}(\q,0)}\ket{\w_n}$ and  $ \ket{w_n} = (\cdots,
\w_n, \cdots)^T$ (see appendix \ref{sec:appendix-A} for more details). 
This is the
DMFT-analog to the Eliashberg equation in the standard theory of 
superconductivity and was first used by Jarrell and co-workers
\cite{JarrellPangCox97,Jarrell95}.

\section{Results for the Phase Boundary}
\label{sec:phase-boundary}

As discussed before, all two-particle \gf s are evaluated within DMFT
in the paramagnetic phase using standard procedures
\cite{AndersJarCox97,Anders99}. The temperature $T_c$ at which the
pair susceptibility diverges defines the boundary between the
paramagnetic and the superconducting phase, if it is the highest
instability temperature.

\begin{figure}[t]
  \begin{center}
\includegraphics[width=85mm]{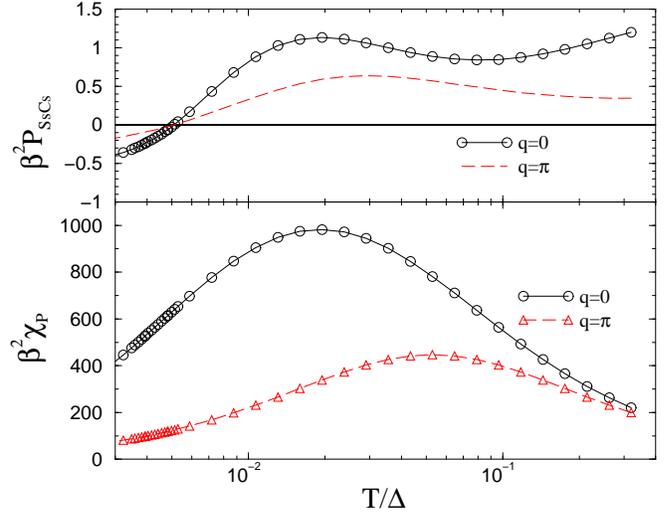}

\caption{Contribution $P^{SsCs}$ to the pair \sus\ $\chi_{P}^{SsCs}(\q \, )$
  from the dominant maximal eigenvalue $|\lambda|$ vs temperature
  in the singlet/singlet (SsCs)  sector for  $\q=0$ and $\q=\pi(1,1,1)$ (upper
  plot). The lower plot shows the total pair 
  \sus\ $\chi_{P}$ given by Eqn
  (\ref{equ-pair-susceptibility}).  Parameters: $\e_{\as} = -3$ and $n_c =2$.
}
\label{fig:pair-sus-CsSs}
\end{center}
\end{figure}

\subsection{Sector I: Spin Singlet/Channel Singlet (SsCs)}

The pair operator in the odd-frequency SsCs sector 
\begin{eqnarray}
  \label{eq:pair-deriv-ss}
 P^{odd}_{ss}(\q,\ivn)  &= & \frac{2}{\beta
   N}\sum_{\k\sigma\,\iwn} \sigma \; \iwn S(\k) 
\\
&&
\times
c_{\k\sigma+}(\iwn +\ivn)
 c_{-\k+\q-\sigma-}(-\iwn)   
\nonumber
\end{eqnarray}
is derived from (\ref{eq:pair-ss}), where $S(\k) = 1$ for isotropic
and $S(\k) = \e_{\k}/V$ of {\em ``extended s-wave''} pairs. Hence,
the pair \sus\ $\chi_{P}^{ss}$ consists of 
a spin diagonal and a spin or channel flip contribution. 
Now, the spin and channel indices, which have been neglected in section
\ref{sec:pp-gh} in order to focus on the structure rather than on details,
are of importance: the resulting coupled
Bethe-Salpeter equations 
yield
\begin{eqnarray}
   \chi_{P}^{SsCs}(\q \, ) &= &-\frac{8}{\beta^2}\sum_{\iwn\iwm}
\iwn\iwm [ \Pi^{dir}(\iwn,\iwm;\q,0)
\non
  \label{eq:pair-ss-pp}
&& - \Pi^{ex}(\iwn,\iwm;\q,0)
]
\end{eqnarray}
in the absence of a magnetic field and  isotropic Cooper-pairs.
$\Pi^{dir}$ describes spin and channel diagonal \pp\ propagation;
$\Pi^{ex}$ the exchange of spin or channel index.
The coupled equations are solved for the \pp\ \gf\ $\Pi^{ss}
= \Pi^{dir}- \Pi^{ex}$ from which the effective pp-vertex $ \mat{\Gamma}^{ss}$
entering (\ref{equ-pair-susceptibility}-\ref{eq:def-mat-M})
\begin{eqnarray}
  \label{eq:eff-pp-vertex-ss}
  \frac{1}{\beta} \mat{\Gamma}^{ss} & = &
\left[\mat{\chi}_{loc}\right]^{-1}
-
\beta
\left[\mat{\Pi}^{dir}_{loc}- \mat{\Pi}^{ex}_{loc}\right]^{-1}
  \\
\chi_{loc}(\iwn,\iwm) &=& \delta_{n,m}G_c(\iwn)G_c(-\iwn)
\end{eqnarray}
is obtained for the singlet/singlet sector, where $G_c$ denotes the
conduction electron \gf.

Jarrell \etal\ investigated the
pair \sus\ for the two-channel Kondo model in the strong
coupling limit using QMC \cite{JarrellPangCox97}. They did not find a maximum
positive eigenvalue $\lambda_m$ with $\lambda_m(T_c) = 1$ for a given
finite temperature for $\q=0$ or $\q=\pi(1,1,1)$  which is equivalent to $\eta(\q)=-1$ (staggered
pairing). However, they observed that 
a single eigenvalue  $\lambda$ approaches $-\infty$. 
 By lowering the temperature further, the associated eigenvector,
 i.~e.\ the
 symmetry form-factor, corresponds to a very large positive eigenvalue
which decreases with decreasing temperature leading to a sign
change of the contribution $P_{CsSc}$ to $\chi_{P}^{SsCs}$ (see Fig.~3
of Ref.~\cite{JarrellPangCox97}).  Since an eigenvalue $\lambda > 1$ was
found,  it was interpreted  as an
indicator for a possible  superconducting instability using a
linearized Eliashberg equation. 
The simultaneous sign change of the staggered and the $\q=0$
contribution is clearly visible in Fig.~3
of Ref.~\cite{JarrellPangCox97} which looks very similar to the upper
plot in Fig.~\ref{fig:pair-sus-CsSs}. Since $\lambda$ appears to jump
suddenly to a value $\lambda>1$, it has been interpreted as a
locally driven first order phase transition \cite{JarrellPangCox97}. 
The \sus\ corresponding to the composite order
parameter $\chi_{P}^{SsCs}$, which would  indicate a true instability of the
whole system, was not investigated.

In general, the local conduction-electron \pp\ \gf s are obtained from a local
two-particle t-matrix equation analogous to the ph-\gf\
(\ref{eq:eq-ph-cc-propagator}) for models with non-interacting
conduction bands.  The local connected part of the two-particle
$T$-matrix  cumulant is given by 
\begin{eqnarray}
  \label{eq:local-t-matrix-ss}
  {\cal T}_{loc}^{ss}(\iwn,\iwm) &=& V^2\left( 
\Pi_f^{dir}(\iwn,\iwm;0) -\Pi_f^{ex}(\iwn,\iwm;0) \right .
\non
&&
\left.  -\beta F(\iwn)
    F(-\iwn)\delta_{n,m}\right) 
\punkt
\end{eqnarray}
Within NCA, the spin or channel exchanging two-particle f-\gf\ $\Pi_f^{ex}$,
depicted in Fig.~\ref{fig:local-pp-prop}, is of the order of
$O(v^0=1)$, where  $v=V\rho$. However, the local  two-particle f-\gf,
diagonal in spin and channel indices, turns out to be of the order
$O(v^2)$ and, therefore, is neglected leading to a total contribution
of $\Pi_f^{pp}(\iwn,\iwm;0)$ 
given by  (\ref{equ-pi-f-ex-nca}). We solved the Eliashberg-like equation
(\ref{eq:base-m})  and confirmed the results of Jarrell \etal\
\cite{JarrellPangCox97} for the two-channel \pam.
The temperature dependence of $P^{SsCs}$, depicted in the upper plot of
Fig.\ \ref{fig:pair-sus-CsSs},
shows excellent agreement with the published results for the two-channel
Kondo model (see Fig.~3 of Ref.\ \cite{JarrellPangCox97}).  

In Appendix \ref{sec:appendix-B}, we show that the sign change of a
contribution to  the pair 
\sus\ is equivalent to finding a non-trivial solution to the
linearized DMFT self-consistency condition (SCC)  in the symmetry broken
phase. This will be discussed later
in Sec \ref{sec:odd-one-part-dmft} in more detail. However, the form 
factor which was found for the Kondo model  \cite{JarrellPangCox97}
and the two-channel \pam\ in our calculation appears to be singular at
$\w=0$: $F(\w) \propto 1/\w$. This has the effect that  a
linearization of the DMFT equations, on which 
the linearized Eliashberg equation is based on, is not justified for
$|\w|\rightarrow 0$, 
and the full SCC has to be considered before any conclusion about the
existence of a symmetry broken state can be drawn for sector
I. Additionally,  the \sus\ of the composite order parameter   
$\chi_{P}^{SsCs}$ is three orders of magnitude larger than the
particular component $P^{SsCs}$ exhibiting the sign change which, therefore,
might be  an irrelevant contribution to the total pair-\sus\ of
the   composite order parameter $O_{ss}$.

We investigated the SsCs sector for a large range  of  parameters
and did not detect a phase transition or a
sign change in  $\chi_{P}^{SsCs}$. The sign change of the contribution
$P^{ScCs}$ to  $\chi_{P}^{SsCs}$ has been viewed as indication of a
first order phase transition by Jarrell and coworkers
\cite{JarrellPangCox97}. However, we always find a higher transition
temperature in  the spin-triplet, channel triplet sector for the
investigated parameters of the two-channel Anderson lattice model. We
believe that the charge fluctuations on the $f$-shell plays a major
role in the formation of the superconducting phase of the Anderson
lattice model as can be seen from Eqns. (\ref{equ-ss-order-parameter})
and (\ref{equ-tt-order-parameter}) while these charge fluctuations are
absent in the Kondo-lattice model.
The superconducting gap function
is dominated by an so-called {\em ``extended s-wave''} ( $\Gamma_1$)
form factor $S(\k) = \sum_{\alpha} \cos(k_\alpha)$ 
parameter in the extreme \IV\ regime, $\Delta E=0$.

\subsection{Sector II (SsCt) and Sector III (StCs)}

We looked into the even frequency eigenspace of $\mat{\Gamma}^{ss}$
(\ref{eq:eff-pp-vertex-ss}) and $\mat{\Gamma}^{tt}_{even}$

\begin{equation}
  \label{eq:eff-pp-vertex-tt-even}
  \frac{1}{\beta} \mat{\Gamma}^{tt}_{even} = 
\mat{E}_{even}\left(\left[\mat{\chi}^{pp}_{loc}\right]^{-1}
-
\beta
\left[\mat{\Pi}^{dir}_{loc}+ \mat{\Pi}^{ex}_{loc}\right]^{-1}
\right)\mat{E}_{even}
\komma
\end{equation}
which belong to the StCs and the SsCt
sector, respectively. 
 $\mat{E}_{even}|_{n,m} = (\delta_{n,m}+\delta_{m,n})/2$ projects
out the odd-frequency components.
No instability was found in either
sector. This was expected from the analytic form of two-particle T-matrix
(\ref{equ-pi-f-ex-nca}) which is asymptotically odd in its frequency
arguments (see also Sec.~\ref{sec:local-two-part-prop}).

\begin{figure}[t]
  \begin{center}
\includegraphics[width=85mm]{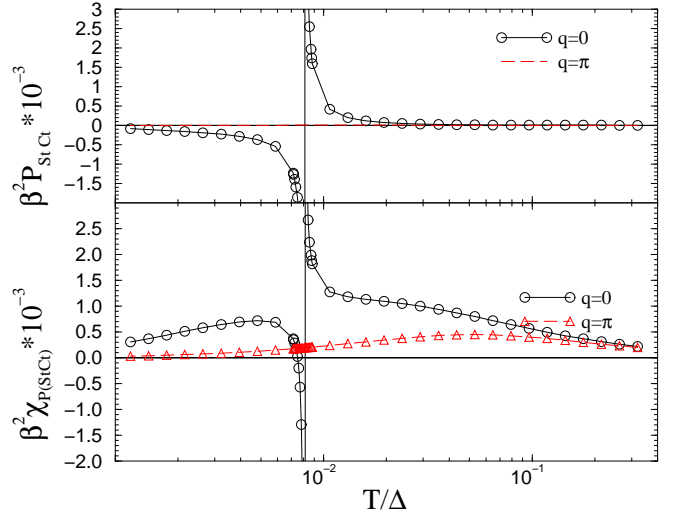}

\caption{Contribution $P^{StCt}$ to the pair-\sus\ $\chi_{P}$ from the
largest positive eigenvalue vs temperature in the triplet/triplet
(StCt)  sector for $\q=\pi(1,1,1)$ (staggered) and $\q=0$ (upper plot), and 
total pair-\sus\ $\chi_{P}$, Eqn.~(\ref{eq:pair-ss-pp}), vs
temperature (lower plot). 
  Parameters: $\e_{\as} = -3$ and $n_c =2$.
}
\label{fig:pair-sus-CtSt}
\end{center}
\end{figure}

\begin{figure}[t]
  \begin{center}
    \begin{tabular}{cc}
\includegraphics[width=75mm]{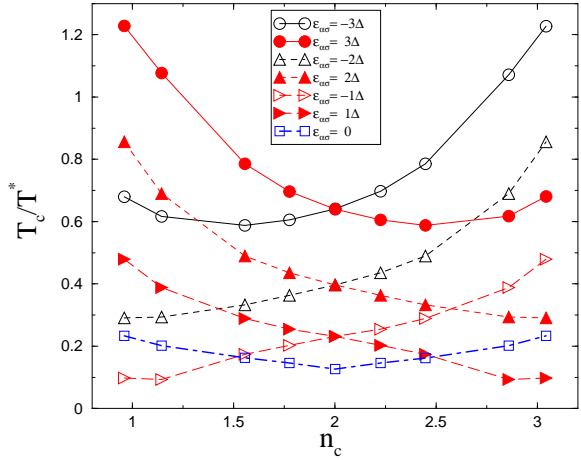}
& (a)
\\
\includegraphics[width=70mm]{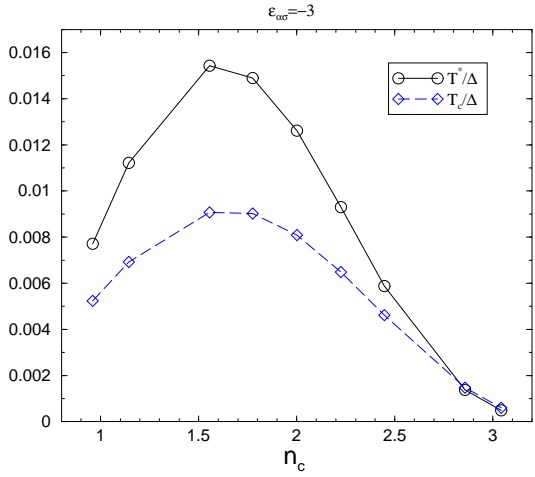}
& (b)
    \end{tabular}

\caption{Reduced transition temperature $t_c = T_c/T^*$ (a) as a function of
  the band filling $n_c$ for  four different values of $|\e_{\as}|$ and
  (b) the dependence of $T_c$ and $T^*$ on  $n_c$ for $\e_{\as} = -3\Delta$.
The components of superconducting order parameter $O_{ij}$ are  invariant under
$\e_{\as}\rightarrow -\e_{\as}$ and a simultaneously performed  particle-hole
transformation, as can be clearly seen in the
plot. Even though $T_c$ increases for decreasing $|\e_{\as}|$, the ratio
$T_c/T^*$ decreases since increasing charge fluctuation tends to
destabilize the superconductivity.
}

\label{fig:pair-sus--vs-nc-CtSt}
\end{center}
\end{figure}

\subsection{Sector IV:  Spin Triplet/Channel Triplet (StCt)}
\label{sec:stct}

The spin triplet, channel triplet  sector, StCt, is characterized by a
tensor order parameter 
with nine components (\ref{equ-tt-order-parameter}) (see also table
\ref{tab:sym-order-parameter}).  
Some of  the  two-particle \gf s in this sector
involve both a direct and an exchange contribution yielding
a matrix representation in Matsubara frequency space which shows an
odd-frequency dependence in order to  obey Pauli's principle, for
example: the spin and channel diagonal \gf. This structure is already
revealed in the free local 
\pp\ \gf 
\begin{eqnarray}
  \label{eq:local-free-prop-tt}
  \chi_{loc}^{\sigma\alpha}(\iwn,\iwm;\ivn) &=&
  G_{\sigma\alpha}(\iwn+\ivn)
G_{\sigma\alpha}(-\iwn)
\non
&&
\times \left(
  \delta_{n,m}
-
  \delta_{-\iwn,\iwm+\ivn}
\right) 
\end{eqnarray}
for $\ivn=0$.
Consequently, the  matrix inversion is mathematically ill-defined in
the particle-particle version 
of Eqn.\ (\ref{eq:irr-vertex-t-matrix}) since the determinant of an
odd matrix in Matsubara frequency space must vanish. This
problem can be circumvented by defining a matrix inversion for
odd-frequency matrices in relation to an odd frequency unitary matrix
$E_{odd}(n,m) = (\delta_{n,m}-\delta_{n,-m})/2$.
This analysis, however,  is only necessary
for some of the nine different pp-\gf s which are needed for the diagonal
order parameter \suss\ $\chi_{P, ij}^{StCt} =
\onepart{O_{ij}}{O^\dagger_{ij}}{0}$. 
By  careful analysis of all triplet two-particle \gf s,
it can be shown that 
all odd-frequency pair-\suss\ are equal in DMFT, and the relevant
irreducible vertex is given by  
\begin{equation}
  \label{eq:eff-pp-vertex-tt}
  \frac{1}{\beta} \mat{\Gamma}^{tt} = 
\mat{E}_{odd}\left(\left[\mat{\chi}^{pp}_{loc}\right]^{-1}
-
\beta
\left[\mat{\Pi}^{dir}_{loc}+ \mat{\Pi}^{ex}_{loc}\right]^{-1}
\right)\mat{E}_{odd}
\komma
\end{equation}
where $\mat{E}_{odd}$ projects out the
odd-frequency components.

For the \tc\ Kondo-lattice model,  $\chi_{P,zz}^{StCt}$  was
investigated by Jarrell \etal\ \cite{JarrellPangCox97}, and no
superconducting instability was found for Kondo couplings of $J/t\ge
0.4$. 
A different picture emerges in the two-channel \pam.
In Fig.~\ref{fig:pair-sus-CtSt}, the total pair-\sus\
$\chi_{P}^{StCt}$ (lower curves)  and the contribution 
stemming from the largest eigenvalue $P^{StCt}$ (upper curves) are
shown in the stable moment regime for half-filling and for the center
of mass  momenta $\q=0$ 
and $\q=\pi(1,1,1)$. A sign change is observed  in both quantities
at a temperature of $T\approx 8\cdot 10^{-3} \Delta$  and for $\q=0$,
which originates from the divergency of  $P_{StCt}$ as
$\lambda_{\mbox{max}}\rightarrow 1$. Consequently, we find a clear
tendency towards a second order phase transition in the
triplet/triplet sector for an uniform order parameter. The pair-\sus\
of a staggered  pair is much smaller than  $\chi_{P}^{StCt}(\q=0)$ and
stays continuous in the vicinity of $T_c$.
The qualitative difference of the  Kondo-lattice model and \pam\ is
the possibility of change fluctuation on the $f$-shell in the latter model,
which is not possible in the first.

The characteristic temperature of the lattice, $T^*$, was
phenomenologically defined as the temperature at which the ratio of
effective local moment $\mu^2_{eff}(T) = T\chi_{loc}(T)$ and the
free moment $\mu^2_{free} = \dps \lim_{T\rightarrow\infty}
\mu^2_{eff}(T)$ equal $0.4$ : $\mu^2_{eff}(T^*)/\mu^2_{free} =
0.4$.  $\chi_{loc}(T)$ is the local \sus\ of the ground state
spin. Therefore, 
$T^*$ is a measure of the Kondo screening of the local magnetic or
quadrupolar spin.

We  investigated the pair-\sus\ $\chi_{P}^{StCt}$ in the \IV, the
Kondo and the stable moment regime for band-fillings 
between quarter and $3/4$-filling and COM momentum of $\q=0$ and
$\q=\pi(1,1,1)$. We always performed a finite size scaling analysis
for the largest eigenvalue of $\mat{M}$,
Eqn.~(\ref{eq:def-mat-M}),  with respect to the matrix size:
$\lambda_m$ approaches a constant for $30-40$ Matsubara
frequencies. The superconductivity is driven by the low energy
components of the two-particle vertex in contrary to the magnetic
phase transitions
\cite{Anders99} where $400-500$ Matsubara frequencies are needed in
order to obtain a  \sus\  invariant of the size of the matrix.
A superconducting  transition  was always obtained, but only
for $\q=0$. Although $T_c$ decreases with decreasing 
coupling constant $g=\rho_0 J=\Delta/|\e_{\as}|$, the normalized
transition temperature $t_c = T_c/T^*$ shown an increase due to the
exponential sensitivity of $T^*$ on $g$ as depicted in
Fig.~\ref{fig:pair-sus--vs-nc-CtSt}(a). 
$T^*$ decreases away from half-filling $n_c>2$: hence $t_c$
increases with $n_c$ as shown in
Fig.~\ref{fig:pair-sus--vs-nc-CtSt}(b). The asymmetry of $T^*$ as
function of $n_c$ simply stems from the asymmetry of energy of the
two local doublets. This asymmetry disappears for $\e_{\as}=0$.

The odd-frequency vertex is attractive (repulsive) in the StCt (SsCs)
sector as seen analytically in the weak coupling limit $V \rightarrow 0$
from (\ref{eq:irr-vertex-dmft}),
(\ref{eq:eff-pp-vertex-ss}) and (\ref{eq:eff-pp-vertex-tt})  in
combination with an NCA approximation for $\Pi_f^{pp}(\iwn,\iwm;0)$ given by 
(\ref{equ-pi-f-ex-nca}). 

\section{Dynamical Mean Field Theory of the Superconducting phase}
\label{sec:dmft-sc}

We have established the phase boundary between the
homogeneous, paramagnetic and the superconducting phase in section
\ref{sec:pp-gh}. Only the one-particle and two-particle  solutions of
the effective site enter the calculation of the pair-susceptibility in
paramagnetic phase. In this section, the modified DMFT equations for
the superconducting phase are derived and approximately evaluated. 
New insight on the physical properties of this phase will be gained.

\subsection{Green Functions}
\label{sec:odd-gfs}

The Greens function  is an $8\times 8$
matrix in the spin and orbital space and formally written as
\begin{equation}
\mat{G}(\iwn,\k) = 
\left(
\begin{array}{cc}
\mat{E}^{-1}(\iwn,\k) & -\mat{\Phi}(\iwn,\k) \\
 -[\mat{\Phi}(-\iwn,\k)]^{h} & \mat{H}^{-1}(\iwn,\k) \\
\end{array}
\right)^{-1} 
\komma
\end{equation}
where $ \mat{H}^{-1}(\iwn,\k) = - \mat{E}^{-1}(-\iwn,\k)$. $\mat{E}$
describes the electron, $\mat{H}$ the hole propagation in the
paramagnetic phase if $\mat{\Phi}=0$ . The relation
between the off-diagonal matrix elements is a consequence of  the definition of the 
\gf\ and, therefore,  is a fundamental relation independent of
possible symmetries of the order parameter.
$\mat{\Phi}(\iwn,\k)$
is a $4\times 4$ matrix in the combined spin/channel space containing
16 linear independent anomalous \gf s. Since the Hamiltonian conserves
magnetic  and channel spin, the natural decomposition is given by 
\begin{eqnarray}
\mat{\Phi}(\iwn,\k) &=& i\mat{\sigma}_2
[ \Phi_0^s(\iwn,\k)\mat{1} +
\vec{\Phi_s}(\iwn,\k)\vec{\mat{\sigma}}] 
\non
&&
\times i\mat{\tau}_2[ \Phi_0^c(\iwn,\k)\mat{1} +
\vec{\Phi_c}(\iwn,\k)\vec{\mat{\tau}}] 
\nonumber
\\
&= & -\mat{\sigma}_2\mat{\tau}_2
\left[
\Phi_0^s(\iwn,\k) \Phi_0^c(\iwn,\k)\mat{1}
\right.
\\
&&
\hspace{13mm}
 + 
\Phi_0^s(\iwn,\k)\mat{1} \vec{\Phi_c}(\iwn,\k)\vec{\mat{\tau}}
\non
&&
\hspace{13mm}
+ \Phi_0^c(\iwn,\k)\mat{1} \vec{\Phi_s}(\iwn,\k)\vec{\mat{\sigma}}
\non
&&
\hspace{13mm}
+
\left.
\vec{\Phi_s}(\iwn,\k)\vec{\mat{\sigma}}\vec{\Phi_c}(\iwn,\k)\vec{\mat{\tau}}
\right]
\nonumber
\komma
\end{eqnarray}
where $\vec{\mat{\tau}}$ acts in channel space and $\vec{\mat{\sigma}}$
in spin  space.
The anomalous components are given by the product \\
$\Phi_0^s(\iwn,\k)\Phi_0^c(\iwn,\k)$ in the SsCs sector,
$\Phi_0^s(\iwn,\k)\vec{\Phi_c}(\iwn,\k)$ in the SsCt sector,
$\Phi_0^c(\iwn,\k)\vec{\Phi_s}(\iwn,\k)$ in the StCs sector
and the tensor $T_{ij}= \Phi_i^s(\iwn,\k)\Phi_j^c(\iwn,\k)$ in the
StCt sector. The similarities to He$^{3}$ are striking: in
He$^{3}$ the order parameter is a tensor which also can be written as a 
product of spin and orbital functions. Focusing on the individual
symmetry sector of interest, the spin/channel matrix
product $\mat{\Phi}(\iwn,\k) [\mat{\Phi}(-\iwn,\k)]^{h}$ is obtained
in the  StCt sector
\begin{eqnarray}
\mat{\Phi}(\iwn,\k) [\mat{\Phi}(-\iwn,\k)]^{h}
&= & 
 \left(\vec{\Phi_c}(\iwn,\k)\vec{\Phi_c^*}(-\iwn,\k)\right)
\\
&&
 \left(\vec{\Phi_s}(\iwn,\k)\vec{\Phi_s^*}(-\iwn,\k)\right)
\mat{1}
\non
& = & \mbox{Tr}\left[\mat{d}(\iwn,\k)\mat{\overline{d}}(-\iwn,\k)\right]\mat{1}
\nonumber
\komma
\end{eqnarray}
where we have introduced the a new parameter matrices 
\begin{eqnarray}
  \label{eq:anomalous-gf-order-param-matrix}
\mat{d}(\iwn,\k)  & =&  \vec{\Phi_s}(\iwn,\k)\vec{\Phi_c}^T(\iwn,\k)
\hspace{5mm} , \hspace{5mm}
\non
\mat{\overline{d}}(\iwn,\k)& =&  \vec{\Phi_c^*}(\iwn,\k)\vec{\Phi_s^*}^T(\iwn,\k)
\end{eqnarray}
and restricted our search to ``unitary states'':
$\vec{\Phi}_{c(s)}(\iwn,\k)\times \vec{\Phi}^*_{c(s)}(-\iwn,\k) = 0$
in obvious analogy to the treatment of He$^{3}$ \cite{WoelfleVollhardt}.
Under the assumption that only the StCt-sector becomes superconducting,
the full $8\times 8$ Greens function is then given by
\begin{eqnarray}
\mat{G}(\iwn,\k) & = &
4
\left(\spur{\mat{E}^{-1}(\iwn,\k)\mat{H}^{-1}(\iwn,\k)}
\right.
\non
&& 
\left.
-4\mbox{Tr}\left[\mat{d}(\iwn,\k)\mat{\overline{d}}(-\iwn,\k)\right]\right)^{-1}
\non
&&
\times \left(
\begin{array}{cc}
\mat{H}^{-1}(\iwn,\k) & \mat{\Phi}(\iwn,\k) \\
 \left[\mat{\Phi}(-\iwn,\k)\right]^{h} & \mat{E}^{-1}(\iwn,\k) \\
\end{array}
\right)^{-1} 
\punkt
\end{eqnarray}
Without directional coupling between spatial and spin/orbit degrees of
freedom, the  tensor $d_{ij}$  separates as
%%$
\begin{equation}
\mat{d}(\iwn,\k) = g(\iwn,\k) \vec{n_s} \cdot \vec{n_c}^T
\komma
\end{equation}
%%$
where $\vec{n_s}$ and $\vec{n_c}^T$ are constant unity vectors in spin and channel
space, and an  amplitude function $g(\iwn,\k) = - g(-\iwn,\k)$.
If the superconducting phase is chosen in such a way that $g$  is
a purely imaginary function of the arguments, and $g^*(-\iwn,\k) =
g(\iwn,\k)$ holds, we obtain the Nambu \gf\ as
\begin{eqnarray}
&&\mat{G}(\iwn,\k)  =  \left[(\iwn-\Sigma(\iwn) -\e_{\k})(\iwn+\Sigma(-\iwn)
+\e_{\k})
\right.
\non
&&\left.
 - g(\iwn,\k)g(\iwn,\k)
\right]^{-1}
\non
&&
%%\cdot 
\times \left(
\begin{array}{cc}
(\iwn+\Sigma(-\iwn)
+\e_{\k})\mat{1} & 
g(\iwn,\k) \mat{\sigma}_2\mat{\tau}_2
\left[
\vec{n_s}\vec{\mat{\sigma}}\vec{n_c}\vec{\mat{\tau}}
\right]
\\
 g(\iwn,\k)\left( \mat{\sigma}_2\mat{\tau}_2
\left[
\vec{n_s}\vec{\mat{\sigma}}\vec{n_c}\vec{\mat{\tau}} 
\right]
\right)^h
&
(\iwn-\Sigma(\iwn) -\e_{\k})\mat{1}
\end{array}
\right)
\; .
\hspace{15mm}
\label{equ:8x8greens-func}
\end{eqnarray}
We have reduced the $8\times 8$
Nambu \gf-matrix  to a standard $2\times 2$ problem for the
amplitude function $g(z,\k)$ and the diagonal self-energy $\Sigma$
with the above assumptions.

\subsubsection{DMFT One-Particle Self-Consistency Condition}
\label{sec:odd-one-part-dmft}

As shown in the previous section, the   $8\times 8$ Nambu \gf\ matrix
can be reduced to a $2\times 2$ matrix  determining
the dynamics of a homogeneous superconducting phase. 
In order to derive  DMFT equations with a purely local
self-energy matrix,  the anomalous
self-energy has to be restricted to isotropic pairs, e.~g.~$S(\k)=1$ 
\begin{eqnarray}
  \label{eq:k-sum-gf}
  \mat{G}_c(z) & =& \frac{1}{N}\sum_{\k} \left[ 
\left[\mat{G}^{0}_{c}(\k,z)\right]^{-1} - \mat{\Sigma}_c(z)
\right]^{-1}
\\
\label{eq:dmft-scc}
\invmat{\mat{\tilde G}_c(z)} & = &\invmat{\mat{G}_c(z)} + \mat{\Sigma}_c(z)
\komma
\end{eqnarray}
where $\mat{\tilde G}_c(z)$ corresponds to the  medium or dynamical mean
field in which the effective impurity is embedded. It is related to a generalized
Anderson width matrix by
\begin{equation}
  \label{eq:anderson-width-matrix}
  \mat{\Delta}(z) = V^2\mat{\sigma}_2 \,\mat{\tilde
    G}_c(z) \,\mat{\sigma}_2 
\end{equation}
and has normal and anomalous components describing quasi-particle
propagation and pair-creation and annihilation, respectively. All
components  enter the calculation of the effective impurity. The
self-energy $\mat{\Sigma}_c(z)$ is determined via
\begin{equation}
\label{eqn:sigma-mat-c}
\mat{\Sigma}_c = \mat{T} \left[\mat{1} + \mat{\tilde G} \mat{T} \right]^{-1}
\komma
\end{equation}
a matrix generalization of the usual DMFT self-consistency equations,
where the diagonal elements of the $T$-matrix are given by the local
quasi-particle scattering matrices $T_e=T_{11}(z)= V^2 G_f(z)$ and
$T_h= T_{22}(z)=  -V^2 G_f(-z)$. The anomalous contribution 
is calculated from
\begin{equation}
  \label{eq:anonalous-t-matrix}
  T_s = T_{12}(\iwn) = - V^4 \frac{1}{\beta} \sum_{\iwm}
  \Pi_f(\iwn,\iwm;0) \tilde f (\iwm)
\punkt
\end{equation}
The negative sign takes into account that
\begin{equation}
\Delta_{12}(z) = -V^2 \tilde G_{12}(z) = -V^2 \tilde f(z)
\punkt
\end{equation}

\subsection{Connection between the Pair-Susceptibility and Green
  Function  in the  Superconducting Phase}
\label{sec:pair-sc-gf}

The DMFT  equations of the previous section are  independent of  the
algorithm to obtain spectral information on the effective site. 
If $g(z)\rightarrow 0$ holds for {\em all} frequencies
when $T\nearrow T_c$, the divergence of the pair-\sus, or
finding an eigenvalue of $\lambda=1$ for the matrix $\mat{M}$, Eqn.\
(\ref{eq:base-m}), is equivalent to solving (\ref{eq:k-sum-gf}) 
and (\ref{eq:dmft-scc}) and
\begin{equation}
\label{equ-linerarized-eliashberg-s}
 g(\iwn)
 =  
\frac{1}{\beta}\sum_{\iwm} \Gamma_c(\iwn,\iwm)\chi_{0}(\q=0,0)
g(\iwm)
\punkt
\end{equation}
is obtained ($\ket{\lambda=1} \propto g(\iwn)$) 
as shown in Appendix \ref{sec:appendix-B}.
As expected in a mean-field theory, the
quasi-particle self-energies are not renormalized. 

However, all eigenvectors with $\lambda=1$ which lead to
superconducting phase transitions reported in Sec.\ \ref{sec:stct}
show a $1/\iwn$ behaviour in leading order reflecting the analytic
properties of the irreducible vertex \cite{CoxZawa98},
Eqn.~(\ref{equ-pi-f-ex-nca}). Therefore, the assumption of a simple
second  order phase transition with $|g(z)|\rightarrow 0$ does not hold,
indicating that also the quasi-particle properties are strongly
renormalized by entering the superconducting phase.
The linearization of Eqns. (\ref{eq:k-sum-gf}-\ref{eqn:sigma-mat-c})
with respect to $g(z)$ is not possible  below $T_c$.
This implies that
Heid's theorem \cite{Heid95}, which states that odd-frequency
superconducting state is associated with a maximum in the free energy,
does not apply to the triplet/triplet superconducting state. Its
its derivation relies on the analyticity  of
$g(z)$ on the real axis.
In general,  $\mat{\Sigma}$ reads
\begin{eqnarray}
\label{eq:sigma-sc}
\mat{\Sigma} & =  &
\frac{1}{A} 
\left(
\begin{array}{cc}
T_e +\tilde h(T_e T_h - T_s^2) & 
T_s - \tilde f (T_e T_h - T_s^2)\\
T_s - \tilde f (T_e T_h - T_s^2) &
T_h +\tilde e(T_e T_h - T_s^2)
\end{array}
\right)
\\
A & = & (1 +\tilde e T_e + \tilde f T_s)
(1 +\tilde h T_h + \tilde f T_s)
\non
&&
-(\tilde e T_s +\tilde f T_h)
(\tilde h T_s +\tilde f T_e)
\komma
\end{eqnarray}
which implies the renormalization of the quasi-particle self-energy
$\mat{\Sigma}_{11}$ by the terms $-\tilde h T_s^2$ even if the we
assume that the quasi-particle scattering t-matrices $T_h$ and $T_e$
are not renormalized close to  $T_c$.

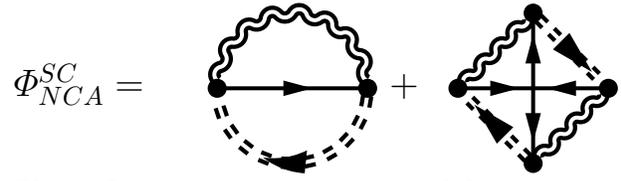
\begin{figure}[tb]

    \begin{picture}(80,25)
\put(0,9){{\Large $\Phi_{NCA}^{SC} =$}}
\put(26,0){
    \begin{fmfchar*}(20,20)
      \fmfpen{0.8thick}
      \fmfstraight
      \fmfleftn{v}{1}
      \fmfrightn{c}{1}
      \fmfmydot{0.1w}{c1}     
      \fmfmydot{0.1w}{v1}     
      \fmf{fermion}{v1,c1}
      \fmf{dbl_wiggly,left=1}{v1,c1}
      \fmf{dbl_dashes_arrow,left=1}{c1,v1}
    \end{fmfchar*}
}
%%%
\put(50,9){{\Large $+$}}
\put(58,0){
    \begin{fmfchar*}(20,20)
     \fmfpen{0.8thick}
      \fmfstraight
      \fmfleftn{l}{1}
      \fmfrightn{r}{1}
      \fmftop{t1}
      \fmfbottom{b1}
      \fmfmydot{0.1w}{r1}     
      \fmfmydot{0.1w}{l1}     
      \fmfmydot{0.1w}{t1}     
      \fmfmydot{0.1w}{b1}     
      \fmf{dbl_wiggly}{l1,t1}
      \fmf{dbl_wiggly}{r1,b1}
      \fmf{dbl_dashes_arrow}{t1,r1}
      \fmf{dbl_dashes_arrow}{b1,l1}
      \fmf{fermion}{l1,c1}
      \fmf{fermion}{r1,c1}
      \fmf{fermion}{v1,b1}
      \fmf{fermion}{v1,t1}
%%      \fmf{plain,left=0.8}{v1,v2}
%%      \fmf{plain}{v1,v2}
%%
    \end{fmfchar*}
}
    \end{picture}

    \caption{Diagrammatic representation of the generating
      functional $\Phi_{NCA}^{SC} $ for the NCA in the superconduction
      (SC)
      phase: the 
      first diagram is the normal contribution, the second  one is the 
      anomalous contribution. The wiggly lines represent the
      \gf\ $P_{\alpha}$ of the quadrupolar doublet, the dashed lines
       $P_\sigma$ for the magnetic doublet
      and the solid lines are the appropriate components of the
      media \gf-matrix {$\mat{\tilde G}$}. 
      The second term  generates contributions to  the
      self-energies of the ionic \gf s and the anomalous
      T-matrix. 
}
    \label{fig:nca-scc-sc}
\end{figure}

\begin{figure}[tb]
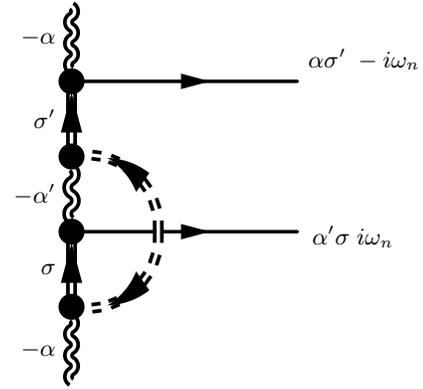

  \begin{center}

    \begin{fmfchar*}(30,50)
      \fmfpen{0.7thick}
      \fmfstraight
      \fmfleftn{v}{6}
      \fmfrightn{c}{6}

      \fmf{dbl_dashes_arrow,right=0.5,tension=0.8}{h1,v4}
      \fmf{dbl_dashes_arrow,left=0.5,tension=0.8}{h1,v2}
       \fmf{phantom}{h1,c3}

      \fmflabel{$\alpha \sigma'\; -\iwn$}{c5}
      \fmflabel{$\alpha'\sigma\; \iwn$}{c3}
      
      \fmf{fermion}{v3,c3}
      \fmf{fermion}{v5,c5}
      
      \fmf{dbl_wiggly,label.side=left,label=$-\alpha$}{v1,v2}
      \fmf{dbl_plain_arrow,label.side=left,label=$\sigma$}{v2,v3}
      \fmf{dbl_wiggly,label.side=left,label=$-\alpha'$}{v3,v4}
      \fmf{dbl_plain_arrow,label.side=left,label=$\sigma'$}{v4,v5}
      \fmf{dbl_wiggly,label.side=left,label=$-\alpha$}{v5,v6}
      \fmfmydot{0.1w}{v2}     
      \fmfmydot{0.1w}{v3}     
      \fmfmydot{0.1w}{v4}     
      \fmfmydot{0.1w}{v5}     
    \end{fmfchar*}

    \caption{Diagrammatic representation of the local anomalous T-matrix 
      in NCA plotted on an imaginary time axis, from bottom to
      top. It was obtained by contracting the incoming lines 
      of the diagram shown in  Fig.\ \ref{fig:local-pp-prop}.
      The double dashed line represents
      anomalous Cooper-pair bath function $\tilde f(z)$ which obeys
      $\tilde f(z) = - \tilde f(-z)$.}
    \label{fig:local-a-tmat}
  \end{center}
%%\end{figure}
\end{figure}

\subsection{Modified Non-Crossing-Approximation for the Superconducting Phase}
\label{sec:mod-nca-sc}

$T_{s}$ is generated by the second term of the generating functional 
for the effective site in a superconducting medium depicted  in
Fig.~\ref{fig:nca-scc-sc}. The diagrammatic
representation of $T_s$ is shown  in Fig.\ \ref{fig:local-a-tmat}.
We have   evaluated  (\ref{eq:anonalous-t-matrix}) in NCA by four integrals along the
branch-cuts by convoluting the Cooper-pair media \gf\ $\tilde f(z)$
with $\Pi_f(\iwn,\iwm;0)$, Eqn. (\ref{equ-pp-f-ex}). 
It was shown that the instability in $\chi_{P}$  brings forth a
solution of (\ref{eq:dmft-scc} - \ref{eq:anonalous-t-matrix}) when expanded
linearly in the anomalous \gf s and medium (see also Appendix \ref{sec:appendix-B}).

Since we cannot use a  linearized  version of the DMFT close to $T_c$
as discussed in the previous section,
we iterated the full DMFT for the superconducting
phase, Eqns (\ref{eq:dmft-scc} - \ref{eq:anonalous-t-matrix}), without 
any further approximations for
$1-T/T_c\ll 1$. The generating functional of the effective site 
 $\Phi_{NCA}^{SC}$ is shown in Fig.\ \ref{fig:nca-scc-sc}.
For $T>T_c$, the equations are identical to the normal state DMFT
since $g(z)$ iterates to zero.

\subsection{One-particle Spectra in the Triplet-Triplet Sector}
\label{sec:tt-gf}

\begin{figure}[htbp]
  \begin{center}
   \includegraphics[width=80mm]{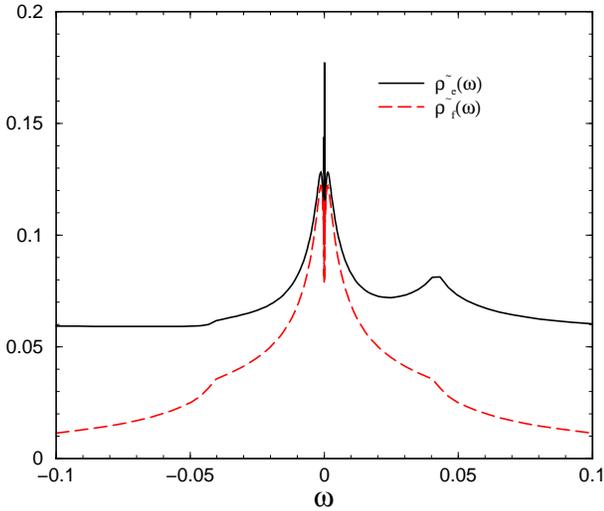} 

    \caption{Quasi-particle media spectral function $\tilde \rho_e(\w)
      = \Im m \, e(\w-i\delta)/\pi$ (solid line) and anomalous media spectral function $\tilde \rho_f(\w)
      = \Im m \, f(\w-i\delta)/\pi$ (dashed line) 
in DMFT(NCA) in the superconducting
      phase close to $T_c$.  $\e_{\as} =-2$.
%%     Parameters as in       Fig. \ref{fig:odd-CtSt-super-gf}. 
}
\label{fig:media-nca-sc}
    
  \end{center}
\end{figure}

\begin{figure}[t]
  \begin{center}
\includegraphics[width=80mm]{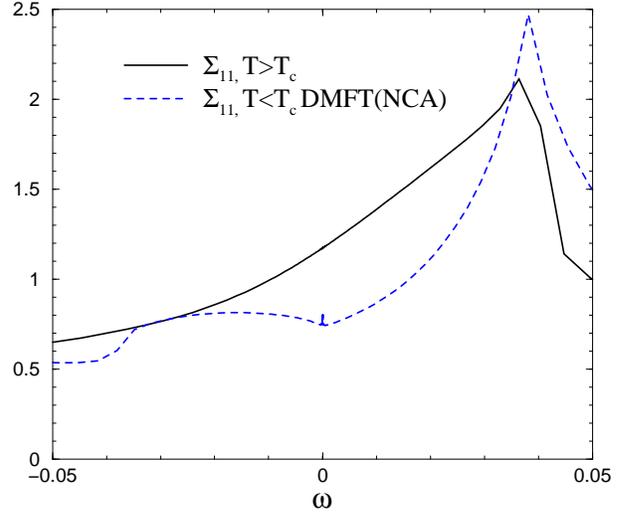}

\caption{Comparison of the imaginary part of the quasi-particle self-energy in the normal
  state, which is also used in  approximation (i), close to $T_c$
  (solid line) and the self-energy $\Im m \Sigma_{11}(\w-i\delta)$
  obtain through (\ref{eq:sigma-sc})  and the full DMFT(NCA) (dashed line).
     Parameters as in       Fig. \ref{fig:media-nca-sc}.
}

\label{fig:odd-CtSt-self-energy}
\end{center}
\end{figure}

\begin{figure}[htbp]
  \begin{center}
    \includegraphics[width=80mm]{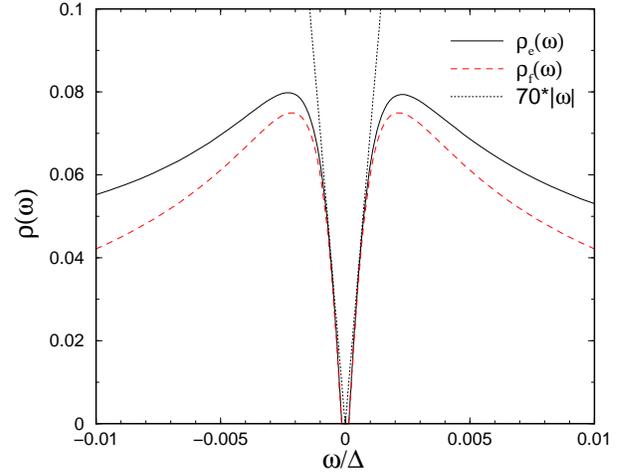}

    \caption{Spectral functions of the quasi-particle and the
      anomalous \gf\  in DMFT(NCA) in the superconducting
      phase for $1-T/T_c\ll 1$ in the vicinity of $\w=0$.    The dotted curve shows a  fit with $a*|\w|$.
      Parameters as in
      Fig. \ref{fig:media-nca-sc}.  
}
\label{fig:spectral-nca-sc}
    
  \end{center}
\end{figure}

We  obtained a self-consistent solution of DMFT(NCA) equations for $T$
close to $T_c$ in the superconducting phase.
For $T<T_c$, the ionic propagators
\cite{Grewe83,Kuramoto83,Bickers87}   aquire an additional self-energy
contribution which originates from the anomalous
conduction-electron bath $\tilde f$ and is generated by the  second
diagram of the generating functional $\Phi_{NCA}^{SC}$ depicted in Fig.\
\ref{fig:nca-scc-sc}.  These
corrections are self-consistently taken into account in our numerics and
modify the quasi-particle $t$-matrices and generate an 
anomalous $t$-matrix as depicted in Fig.~\ref{fig:local-a-tmat}.
We start the iteration with a finite anomalous media  $\tilde f_{0}(z)
= A/(z+iT)$ where the weight $A$ is of the order of $10^{-3}$.
The temperature at which a finite anomalous
t-matrix $T_s$ is stabilized coincides with the instability of the
pair-susceptibility. Above $T_c$,  $\tilde f(z)$ and $T_s(z)$ approach
zero rather rapidly while iterating  Eqns.~(\ref{eq:dmft-scc} - \ref{eq:anonalous-t-matrix}).

The spectral functions of the effective quasi-particle and anomalous media 
is plotted in Fig. \ref{fig:media-nca-sc}: a delta peak at $\w=0$ is
growing on top of the regular part  similar to the reported media  in
the single-channel Anderson lattice at $T=0$, shown in Fig.~3 of Ref.\
\cite{PruschkeBullaJarrell2000}. This  delta peak should lead to a
Fermi-liquid form of the 
quasi-particle self-energy $\Sigma_{11}$ for $T\rightarrow 0$.
 We clearly see the decrease of the quasi-particle
scattering rate slightly below $T_c$ 
as shown in Fig. \ref{fig:odd-CtSt-self-energy}.
This suppression of the
quasi-particle scattering rate indicates the removal of the residual
entropy of the normal state: the magnetic scattering is strongly
reduced due to forming a composite order parameter entangling local
magnetic (quadrupolar) degrees of freedom with spin
singlet/quadrupolar triplet Cooper pairs. 

The local quasi-particle and
anomalous \gf\ is shown in figure \ref{fig:spectral-nca-sc}. It
develops a pseudo-gap  for very low frequencies which behaves like
$|\w|$ as indicated by the dotted fit  curve in Fig.~\ref{fig:spectral-nca-sc}.
This approximate $|\w|$-dependency of the spectral functions can be understood analytically.
It is apparent from the  evaluation of Eqn. (\ref{eq:k-sum-gf}) that
the frequency dependency of the spectral function is determined by 
the imaginary part of the inverse of the determinant
$\sqrt{(z-(\Sigma(z)+\Sigma(-z))/2)^2   - g^2(z)}$. Let us  assume
that the leading contribution to  $g(z)$
can be  approximated by a temperature-broadened Lorentzian
$A^2/(z+i\,\mbox{sign}(\Im m z) \alpha T)$, which reflects the $1/z$
divergency of  $\Pi_f(\iwn,\iwm;0)$ at $T=0$ (see
Eqn.~(\ref{equ-pi-f-ex-nca}));  the weight $A$
is proportional to the order parameter, and $\alpha$ a dimensionless
constant of the order of $O(1)$. Due to this $1/z$ divergency, its
analytic behaviour has similarities to the odd frequency mean-field
solution of the anomalous self-energy by Coleman \etal\
\cite{ColemanMirandaTsvelik94}, who found that $g(z)= \Sigma_{12}
\propto 1/z$.

\section{Conclusion}

There is a controversial debate (see \cite{Heid95} and reference
therein) as to whether an odd-frequency solution
with a COM momentum $Q=0$ is connected to a minimum or a maximum of the
free energy and, hence, may be thermodynamically unstable. 
Assuming a Fermi-liquid normal phase and the
occurrence of an infinitesimal odd-frequency anomalous self-energy
$g(z)$ at $T_c$, which implies a second order phase transition, Heid  
showed  that  one  indeed obtains an increase of the free energy
in the superconducting phase for $Q=0$ \cite{Heid95}. We will give two
arguments 
why we believe, in our case on the contrary, that  the superconducting phase is
thermodynamically stable, even though we cannot prove unambiguously
whether the phase transition is first order or continuous. 
It is straightforward to extend Heid's theorem onto a
non-Fermi liquid normal phase. An infinitesimal and analytic
odd-frequency $g(z)$ does lead to an enhancement of the quasi-particle
DOS $\rho$ at $\w=0$ and hence to a maximum in the free energy. 
However, based on the self-consistent solution for the anomalous \gf s
(\ref{eq:dmft-scc}-\ref{eq:anonalous-t-matrix}),
$g(z)$ converges to a finite value such that $\Gamma_s$ is approximately
$\Gamma_s\approx 
\Gamma_{qp}$ generating a gap in the lattice \gf s  similar to the
one reported earlier \cite{ColemanMirandaTsvelik94}. $g(z)$ might
be approximated in leading order by a Lorentzian with a finite weight
$\tilde g \approx \sqrt{\Gamma_{qp}\alpha T}$. The assumption of an
infinitesimal value of $g(z)$ for all frequencies is violated and
hence Heid's theorem is not applicable in our case.
Since the order parameter then has a finite value at the transition,
the phase transition would be first order and the associated latent
heat $L=T_c(S_N-S_s)$ would be related to the change of entropy.
However, in the presence of an anomalous medium, the quasi-particle
self-energy is also modified and, therefore, the feedback into the effective
site must  be calculated self-consistently. The generating functional 
for the local site in a superconducting medium is shown in
Fig.~\ref{fig:nca-scc-sc} for the NCA. We expect that the 
self-consistent  solution for the one-particle \gf\ in the superconducting phase
will show a vanishing scattering rate for the quasi-particle at
$\dps\lim_{\w\rightarrow 0} \Gamma_{qp}\rightarrow 0$. In this case, a non-analytic
$g(z)$, even with an infinitesimal $\tilde g$,
will be able to always produce a gap.

It turns out, however, that the DMFT(NCA)
equations for the superconducting  phase are
numerically unstable for $T\ll T_c$. We believe that the instability
of the equations is  related to the known inability of the NCA to
reproduce the Fermi-liquid phase of the periodic Anderson model
\cite{Kim87,Kim90}. It is clearly related to the occurrence of an
additional very sharp delta peak in the media spectra (Fig.\ref{fig:media-nca-sc}).
Nevertheless, the solutions obtained at $T_c$ show a  reduction of the local free
energy in the presence of a finite anomalous \gf\ and a tendency
to reduce the scattering rate for the quasi-particles. 
Therefore, even
though we cannot derive conclusively the type of the phase transition, we
believe that there are strong hints towards a second order
transition in the StCt sector. This phase transition should be  associated
with a minimum of the free energy since an energy gap  develops in the
spectrum of the local band \gf\ indicating an energy gain by condensation.

\begin{figure}[ht]
\includegraphics[width=80mm]{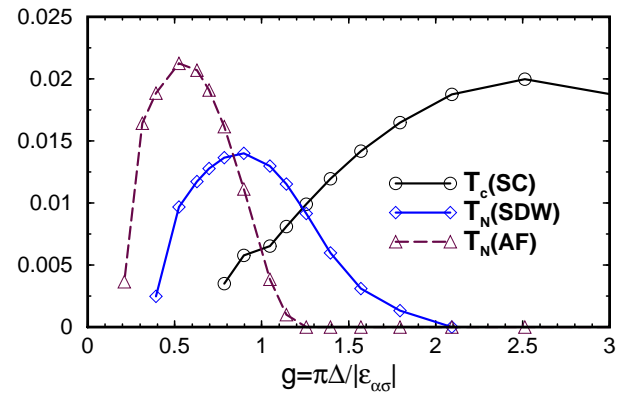}
\caption{The phase diagram of the two-channel periodic Anderson model for $n_c =2.2$:
  the superconducting transition temperature $T_c$, the
  antiferromagnetic  transition temperature $T_m(AF)$, the
  spin-density wave transition  temperature $T_m(SDW)$  are plotted
  versus the effective dimensionless   coupling constant
  $g=\pi\Delta/|\eas|$. Spin/channel
  triplet superconductivity dominates the IV regime  and the cross-over
  region to the Kondo regime. 
}
\label{fig:phase-diagram-magnetic-sc}

\end{figure}

\subsection{Phase Diagram and Experimental Relevance}

The phase diagram shown for $n_c=2.2$ 
in Fig.~\ref{fig:phase-diagram-magnetic-sc} 
summarizes our investigation of superconductivity and
magnetism in the two-channel Anderson model. Superconductivity
dominates the \IV\ regime and the corresponding order parameter has StCt
symmetry. An instability in the SsCs reported in the two-channel Kondo lattice
could be reproduced but occurs at much lower temperatures.

The magnetic or quadrupolar phases of the model was discussed in a
previous publication \cite{Anders99}:
Ferromagnetism is only found
for low band fillings $n_c\approx 1$. Spin-density wave phase
transitions take over in the Kondo-regime at $g\approx 1.3$
corresponding to $\eas=-2.4$. The SDW wave-vector is continuously shifted
towards nearest-neighbour antiferromagnetism, which is suppressed for
$g\rightarrow 0$. Since all calculations were performed in the
paramagnetic phase of the model, the highest transition temperature
defines the nature of the incipient order. It cannot be ruled out, however,
that an SDW phase is replaced by, or even coexists with, a
superconducting phase as found in some Uranium based \HF\ compounds.
Since the quasi-particle spectral function vanished linearly for
$\w\rightarrow 0$, $\rho(w) \propto |\w|$ as depicted in Fig.\
\ref{fig:spectral-nca-sc}, for a spatially isotropic gap function,
 the specific heat obeys a power law $T^\alpha$ below $T_c$. Additional
 anisotropic components $\propto \e_{\k}$ contribute in the \IV\
 regime. Therefore, a exponent $2<\alpha <3$ is expected: a value of
 $\alpha\approx 2.8$ was  experimentally observed on UBe$_{13}$.

In the traditional approaches to Heavy Fermion
superconductivity, one solely focuses on possible $\k$-dependency of
the gap function with nodes on the Fermi surface in order to explain
the power-law dependency of the specific heat, for instance. Here, we
presented a pairing mechanism which leads 
to a gap function with nodes in frequency space additionally to a $\k$
dependence $\propto \e_{\k}$. The pairing interaction correlates
spatially extended, nearest neighbor Cooper-pairs to local spin or
quadrupolar moments. It was not necessary to introduce finite center
of mass momentum Cooper-pairs, i.~e.~$\Q\not=0$, to observe the phase
transition into this state.  Finite $\Q$ Cooper-pairs would imply a
persisting current, and   phases of this kind have sofar not been observed
experimentally.

\begin{figure}[ht]
\includegraphics[width=80mm]{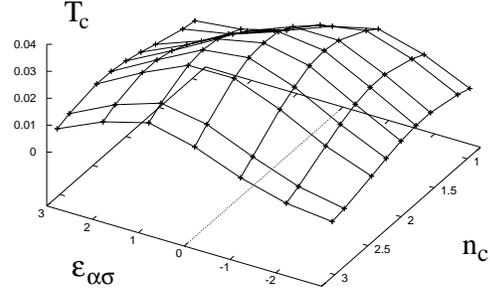}
\caption{The transition temperature $T_c$ as function of the level
  difference $\e_{\as}$ and the conduction band filling. $n_c=2$
  corresponds to half-filling. Note the inversion of the axis scale in
  order to visualize more clearly the dependency of $T_c$ for
  $\e_{\as}\approx 1$ and for large band fillings $n_c$, a regime
  which might be of interest with respect to UBe$_{13}$.
}
\label{fig:tc-vs-ef-nc}

\end{figure}

Assuming a quadrupolar ground state of the Uranium ion,
Fig.~\ref{fig:phase-diagram-magnetic-sc} will show the phase diagram
for quadrupolar order since spin and channel indices are just
exchanged. The superconducting order parameter, however, is not
affected by this transformation as discussed in
Sec. \ref{sec:composite-orderparameter}. 
The  tensor order parameter $O_{ij}$,
(\ref{equ-tt-order-parameter}), transforms as $SO(3)\otimes SO(3)$.
In a cubic crystalline field environment, this order
parameter reduces to $\Gamma_1 \oplus \Gamma_3 \oplus \Gamma_4 \oplus
\Gamma_5$ and the phenomenological 
approach by Sigrist and Rice \cite{SigristRice89} is still applicable in our case.

 Complicated phase  
diagrams as found in Thorium doped U$_{1-x}$Th$_x$Be$_{13}$ are 
explained by order parameters of different
irreducible representation. Recently, the second phase
transition of U$_{1-x}$Th$_x$Be$_{13}$ was  interpreted as a magnetic
phase \cite{Kromer2000} coexisting with the superconducting phase. If
our model has any relevance for U$_{1-x}$Th$_x$Be$_{13}$, we would
expect an antiquadrupolar order or a quadrupolar density wave, since a
magnetic ground state of Uranium contradicts the  weak
magnetic response of van-Vleck type \cite{ub13-chi1}. 

The superconducting transition temperature $T_c(\e_{\as},n_c)$, which is
shown normalized to $T^*$ in Fig.~\ref{fig:pair-sus--vs-nc-CtSt},
is only symmetric with respect to $\e_{\as}$ at half filling. 
We have summarized all the information on 
$T_c(\e_{\as},n_c)/\Delta$ in a 3d-plot displayed 
in Fig.~\ref{fig:tc-vs-ef-nc}.  We note that for $\e_{\as}\approx 1$, i.~e.~an
\IV\ regime with quadrupolar ground state on the local shell as in
Uranium in cubic environment, $T_c$
decreases with increasing hybridization and band filling. A similar 
behaviour is observed in  U$_{1-x}$Th$_x$Be$_{13}$ where $T_c$
decreases under applied pressure and also under Thorium doping which
is interpreted as increasing the number of band-electrons
\cite{Steglich2002privat}.

The skutterudite materials 
PrFe$_4$P$_{12}$ and PrOs$_4$Sb$_{12}$ are also prominent candidates
for the $SU(2) \otimes SU(2)$ lattice model, since Pr$^{3+}$ is hosted  in a
cubic crystalline environment. The anti-quadrupolar
ordered ground state is observed in PrFe$_4$P$_{12}$, while
PrOs$_4$Sb$_{12}$ shows superconductivity below 2K. We expect that
PrFe$_4$P$_{12}$ becomes superconducting under pressure.

\paragraph*{Acknowledgment.}
We would like to thank D.~Cox, P. Gegenwart, N.~Grewe, M.\ Jarrell,
M. Lang and F.~Steglich for many stimulating
discussions. This 
work was funded in parts by an DFG grant AN 275/2-1.

\appendix

\section{Composite Order Parameter}
\label{sec:appendix-order-parameter}

The composite order parameter can be related to the time derivative of
the isotropic and {\em extended s-wave} anomalous \gf s in the
appropriate symmetry sector - see table \ref{tab:sym-order-parameter}
on page \pageref{tab:sym-order-parameter}.
The analytically continued Matsubara \gf\ for fermionic operators $A$
and $B$ obeys the equations of 
motion 
\begin{eqnarray}
\label{equ-eom-ha}
z \; \onepartz{A}{B} & = & \expect{ \{A,B\}} - \onepartz{[H,A]}{B} 
\punkt
\end{eqnarray}
The commutator of a conduction electron  with the local
quantum numbers $\as$ at site $\nu$,
\begin{equation}
[H,c_{\nu\alpha\sigma} ] = -\psi_{\nu\alpha\sigma} -V_\alpha
X_{-\alpha,\sigma}^{\nu}
\komma
\end{equation}
is derived from the \hamil\ (\ref{eq:tca-97}),  where
\begin{eqnarray}
c_{\nu\alpha\sigma} & = & \frac{1}{\sqrt{N}}
\sum_{\k}e^{i\k\vec{R}_\nu} c_{\k\alpha\sigma} 
\\
\psi_{\nu\alpha\sigma} & = & \frac{1}{\sqrt{N}} \sum_{\k}
e^{i\k\vec{R}_\nu} \e_{\k\as}
c_{\k\alpha\sigma} 
\end{eqnarray}
has been used. 
Using the equation of motion 
\begin{equation}
z\; \onepart{c_{\nu\alpha\sigma}}{c_{\nu\alpha'\sigma'}}{z}
=
\onepart{\psi_{\nu\alpha\sigma}}{c_{\nu\alpha'\sigma'}}{z}
+
 V_\alpha
\onepart{X^{\nu}_{-\alpha,\sigma}}{c_{\nu\alpha'\sigma'}}{z}
\label{eqn-cc-xc-I}  
\end{equation}
to calculate the equal time expectation value of the time derivative
$\expect{\frac{d}{d\tau}c_{\nu\as}(\tau) c_{\nu\alpha'\sigma`}}$ in
the form
\begin{equation}
D_{\sigma\sigma`}^{\alpha\alpha'}
= \frac{1}{\beta}\sum_{\iwn} e^{\iwn\delta} 
\iwn
\onepart{c_{\nu\alpha\sigma}}{c_{\nu\alpha'\sigma'}}{\iwn}
\end{equation}
yields
\begin{eqnarray}
D_{\sigma\sigma`}^{\alpha\alpha'} & = &
\expect{\psi_{\nu\alpha\sigma}c_{\nu\alpha'\sigma'}}
+ V_\alpha
\expect{X^{\nu}_{-\alpha,\sigma}c_{\nu\alpha'\sigma'}}
\punkt
\end{eqnarray}
The first term on the r.h.s.~represents the averaged kinetic energy, which
is zero in the case of an odd-frequency anomalous \gf\ 
$\onepart{c_{\k\alpha'\sigma'}}{c_{-\k\alpha\sigma}}{z}$, while
the second describes the coupling to the localized $f$-shell.
Using the commutator 
\begin{eqnarray}
[H, X_{\alpha,\sigma}^\nu]
 &= &
(E_\alpha - E_\sigma)  X_{\alpha,\sigma}^\nu 
-V_{-\alpha}\sum_{\sigma'}  X_{\sigma',\sigma}c_{\nu-\alpha\sigma'}
\non
&&
-\sum_{\alpha'} V_{\alpha} X_{\alpha,-\alpha'}c_{\nu\alpha'\sigma}
\komma \hspace{10mm}
\label{equ-comutator-x}
\end{eqnarray}
and
\begin{eqnarray}
C_{\sigma\sigma'}^{\alpha\alpha'}
= 
\frac{1}{\beta}\sum_{\iwn} e^{\iwn\delta} 
\iwn\left(
\onepart{X^\nu_{\alpha,\sigma}}{c_{\nu'\alpha'\sigma'}}{\iwn}
+
\right.
\\
\left.
\onepart{c_{\nu'\alpha'\sigma'}}{X^\nu_{\alpha,\sigma}}{\iwn}
\right)
\nonumber
\end{eqnarray}
we derive 
%%%%%%%%%%%%%%%%%%%%%%%
%%  begin
%%%%%%%%%%%%%%%%%%%%%%%%
\begin{eqnarray}
\label{equ-define-b}
2V B_{\sigma\sigma'}^{\alpha\alpha'} (\nu)
& =  & C_{\sigma\sigma'}^{\alpha\alpha'}
 -
2(E_\sigma-E_\alpha)
\expect{X^\nu_{\alpha,\sigma}c_{\nu'\alpha'\sigma'}}
\end{eqnarray}
via equation of motion  and, therefore, with $\Delta E = E_\sigma-E_\alpha$ and  $V_\alpha =
\alpha V$
\begin{equation}
|\Delta E|
D_{\sigma\sigma'}^{\alpha\alpha'} - \mbox{sign}(\Delta E) 
\alpha \frac{V C_{\sigma\sigma'}^{\alpha\alpha'}}{2} = -
\alpha V^2 \mbox{sign}(\Delta E) B_{\sigma\sigma'}^{\alpha\alpha'} (\nu) 
\punkt
\label{eqn:d-cprime}
\end{equation}
%%%%%%%%%%%%%%%%%%%%%%%
%% end
%%%%%%%%%%%%%%%%%%%%%%%%
The quantity $B_{\sigma\sigma'}^{-\alpha\alpha'}(\nu)$ is
defined as
\begin{eqnarray}
B_{\sigma\sigma'}^{-\alpha\alpha'}(\nu) &=&
\alpha\sum_{\sigma''}
\expect{ X_{\sigma'',\sigma}c_{\nu\alpha\sigma''}c_{\nu'\alpha'\sigma'}}
\non
&&+
\sum_{\alpha''} \alpha''
\expect{X_{-\alpha,-\alpha''}c_{\nu\alpha''\sigma}c_{\nu'\alpha'\sigma'}}
\punkt
\end{eqnarray}
A  relation  for $ C_{\sigma\sigma'}^{\alpha\alpha'}$ can be
derived using (\ref{equ-eom-ha}) and (\ref{equ-comutator-x})
\begin{equation}
\label{eq:c-aa-t-aa}
\alpha V  C_{\sigma\sigma'}^{-\alpha\alpha'} =
-
2\frac{1}{\beta}\sum_{\iwn} e^{\iwn\delta} 
\iwn
\onepart{\psi_{\nu\alpha\sigma}}{c_{\nu\alpha'\sigma'}}{\iwn}
 = -2 \tilde T_{\sigma\sigma'}^{\alpha\alpha'}
\nonumber
\komma
\end{equation}
and the assumption that the anomalous \gf\ is odd in frequency. If only
pairs with COM momentum $\vec Q =0$ condensate, the condition
$$
\onepart{c_{\k\alpha'\sigma'}}{c_{-\kk\alpha\sigma}}{z} = 
\delta_{\k,\kk}
\onepart{c_{\k\alpha'\sigma'}}{c_{-\k\alpha\sigma}}{z}
$$
holds. Then  $\tilde T_{\sigma\sigma'}^{\alpha\alpha'}$ is equal to
$T_{\sigma\sigma'}^{\alpha\alpha'}$, defined in
Eqn.~(\ref{eq:extended-s-wave}). 
It is straightforward to show that
\begin{equation}
 \left(
\expect{\vec{S}\; \vec{P}^{t,s}} 
- \expect{\vec{\tau}\;  \vec{P}^{s,t}}
\right)
= -\sum_{\alpha\sigma} \sigma B_{\sigma-\sigma}^{\alpha\alpha}
\label{eqn:order-param-app-b}
\end{equation}
and to obtain the equation (\ref{equ-ss-order-parameter}) from
(\ref{eqn:d-cprime}) and (\ref{eqn:order-param-app-b}): 
\begin{eqnarray}
 O_{ss} &= & 
\frac{|\Delta E|}{|V|^2} \sum_{\as} \as D_{\sigma-\sigma}^{\alpha-\alpha}
+\frac{ \mbox{sign}(\Delta E)}{V^2} 
\sum_{\as} \as  T_{\sigma-\sigma}^{\alpha-\alpha}
\punkt
\label{equ-ss-order-parameter-appendix}
\end{eqnarray}
Eqns.~(\ref{eqn:d-cprime}) and (\ref{eqn:order-param-app-b}) in
conjunction also yield  Eqn.~(\ref{equ-tt-order-parameter}) on page
\pageref{equ-tt-order-parameter} for the
order parameters in section IV.

\section{Calculation of the Odd-Frequency Pair-\sus}
\label{sec:appendix-A}

Assuming we know the irreducible \pp\ vertex
$\mat{\Gamma}_{irr}^{pp}$ for the appropriate
symmetry and only consider isotropic pairs,
then  the pair-\sus\ of the composite order parameter is proportional to
\begin{equation}
\label{equ-pair-susceptibility-app} 
\chi_{P}(\q) = 
 \frac{1}{\beta}\sum_{n,n'} \w_n \w_{n'}
\left[\mat{\chi}(\q,0)
 \frac{1}{\mat{1} - \frac{1}{\beta}
\mat{\Gamma}_{irr}^{pp}(0)\mat{\chi}(\q,0)}
\right]_{\iwn\iwnn}
\punkt
\end{equation}
Since $\mat{\chi}(\q,\ivn)$ is
diagonal in Matsubara frequency space, $\sqrt{\mat{\chi}(\q,0)}$ is 
a well defined quantity so that a new symmetrical matrix
$\mat{M}$  
\begin{equation}
  \label{eq:def-mat-M-appendix}
  \mat{M} = \sqrt{\mat{\chi}(\q,0)}
\frac{1}{\beta}\mat{\Gamma}_{irr}^{pp}(0)
\sqrt{\mat{\chi}(\q,0)}
\komma
\end{equation}
can be introduced. It is diagonalized  to obtain the base
$\ket{\lambda}$, Eqn (\ref{eq:base-m}),
$
%\begin{equation}
%  \label{eq:base-m-A}
  \mat{M} \ket{\lambda} = \lambda \ket{\lambda}
%%\punkt
%%
%%
%%\end{equation}
$
and to rewrite $\chi_{P}(\q)$
\begin{equation}
  \label{eq:chi-p-m-matrix}
  \chi_{P}(\q) = \frac{1}{\beta} \bra{\w_n} \sqrt{\mat{\chi}(\q,0)}
\left[\mat{1} -\mat{M}\right]^{-1}  \sqrt{\mat{\chi}(\q,0)} \ket{w_n}
\komma
\end{equation}
where $ \ket{w_n} = (\cdots, \w_n, \cdots)^T$.
This eigenvalue equation is viewed as the analog to the linearized
Eliashberg equation in the standard theory of superconductivity, even
though no reference to Migdal's theorem was made here.
We can use all eigenvectors $\ket{\lambda}$ as a complete base in
frequency space
\begin{equation}
  \label{eq:lambda-m-base}
  \hat 1 = \sum_{\lambda} \ket{\lambda}\bra{\lambda}
\end{equation}
and insert it into (\ref{eq:chi-p-m-matrix}) to obtain
\begin{equation}
  \label{eq:chi-p-contri}
  \chi_{P}(\q \, ) =  \frac{1}{\beta}\sum_{\lambda} 
| \underbrace{\bra{\lambda}
  \sqrt{\mat{\chi}(\q,0)}\ket{\w_n}}_{c_\lambda}
|^2\frac{1}{1-\lambda} 
\punkt
\end{equation}
Since $\sqrt{\mat{\chi}(\q,\ivn)}\ket{\w_n}$ is an odd-frequency
vector, only odd-frequency eigenvectors $\ket{\lambda}$ contribute
to $ \chi_{P}(\q)$. The leading contribution is given by the
eigenvector $\ket{\lambda_c}$ for which $|1-\lambda|$ has a
minimum. Hence, we can write
\begin{equation}
  \label{eq:pair-sus-lambda-appendix}
   \chi_{P}(\q) = \underbrace{\frac{1}{\beta} c_{\lambda_c}^2
     \frac{1}{1-\lambda_c} }_{P_{s}(\q)} 
+ \mbox{regular terms}
\end{equation}
It is now obvious that the pair-\sus\ $\chi_{P}(\q)$ diverges when 
$\lambda_c\rightarrow 1$, and that  $P_{s}(\q)$ is positive in the paramagnetic
phase.

\section{One-Particle Self-Consistency Condition for Local Cooper Pairs}
\label{sec:appendix-B}

In DMFT, a purely local
self-energy leads to a  self-con\-si\-stency condition (SCC), which states
that the $\k$-summed one-particle lattice \gf\ equals the \gf\
calculated for  an effective site with excitations into a surrounding
effective medium:
\begin{eqnarray}
  \label{eq-appB:k-sum-gf}
  \mat{G}_c(z) & =& \frac{1}{N}\sum_{\k} \left[ 
\left[\mat{G}^{0}_{c}(\k,z)\right]^{-1} - \mat{\Sigma}_c(z)
\right]^{-1}
\\
\label{eq-appB:dmft-scc}
\invmat{\mat{\tilde G}_c(z)} & = &\invmat{\mat{G}_c(z)} + \mat{\Sigma}_c(z)
\punkt
\end{eqnarray}
$\mat{G}^{0}_{c}(\k,z)$ denotes the \gf\ in the absence of any
interaction, and
$\mat{\tilde G}_c(z)$ denotes the  medium or dynamical mean
field which enters the effective site problem. It is related to the
Anderson width matrix by
\begin{equation}
  \label{eq-appB:anderson-width-matrix}
  \mat{\Delta}(z) = V^2\mat{\tilde \sigma}_2\; \mat{\tilde
    G}_c(z)\mat{\tilde \sigma}_2 
\end{equation}
where $\mat{\tilde \sigma}_2 $ is the $8\times 8$ generalization of
$\sigma_2$ appropriate to the description of the superconducting state
with spin and channel degrees of freedom. It was assumed in
Sec.~\ref{sec:odd-gfs} that the 
orientation in spin and channel space is described by the two 
unitary vectors. Hence, we use a $2\times 2$ matrix
formalism for the dynamics. Close to
$T_c$, e.g. for $g(z)\rightarrow 0$, Eqn.~(\ref{equ:8x8greens-func}) is
approximated by
\begin{eqnarray}
\mat{G}_c(z,\k) & = & \frac{1}{(z-\Sigma_e -\e_{\k})(z-\Sigma_h
+\e_{\k})}
\\
&&
\times
\left(
\begin{array}{cc}
z-\Sigma_h +\e_{\k} &  g(z)
\\
 g(z)& z-\Sigma_e -\e_{\k}
\end{array}
\right)
\hspace{10mm}
\label{equ:2x2greens-func}
\non
\Sigma_e & = & \Sigma(z) \hspace{5mm} ;  \hspace{5mm}
\Sigma_h =  -\Sigma(-z) 
\end{eqnarray}
in linear order in $g(z)$. The subscripts $e$ and $h$ refer to
electrons or holes. We immediately recognize, in conjunction with
Eqn.~(\ref{eq-appB:k-sum-gf}), that the quasi-particle properties are not
renormalized close to $T_c$. This is typical for a mean-field theory
as the DMFT. 
With the local electron and hole \gf s
\begin{eqnarray}
  \label{eq-appB:e-qp}
  e(z) & = & \frac{1}{N}\sum_{\k} \frac{1}{z-\Sigma_e -\e_{\k}}
\komma
\\
%%\hspace{2mm} \mbox{and}  \hspace{2mm}
  h(z)  &=&  \frac{1}{N}\sum_{\k} \frac{1}{z-\Sigma_h +\e_{\k}} = -e(-z)
\komma
  \label{eq-appB:h-qp}
\end{eqnarray}
we obtain the off-diagonal matrix element
\begin{equation}
  \label{eq-appB:define-f}
  f(z) = \left[\mat{G}_c(z)\right]_{12} = \frac{e(z) + h(z)}{2z-\Sigma_e
    -\Sigma_h} g(z)
\komma
\end{equation}
which we insert into the SCC condition (\ref{eq-appB:dmft-scc}) with
the result
\begin{equation}
  \label{eq-appB:off-media}
  \frac{\tilde f(z)}{\tilde e(z) \tilde h(z)} 
= 
  \frac{ f(z)}{ e(z)  h(z)}  - g(z)
\komma
\end{equation}
where $\tilde e = \left[\mat{\tilde G}_c(z)\right]_{11}$  corresponds
to the electron medium, and $\tilde h= \left[\mat{\tilde
G}_c(z)\right]_{22}$ corresponds to the 
hole medium. $\tilde f =  \left[\mat{\tilde G}_c(z)\right]_{12}$
denotes the anomalous medium in the superconducting phase. 
$g(\iwn)$
is generated by the particle-particle irreducible vertex
$\Gamma_c^{pp}(\iwn,\iwm)$ 
\begin{equation}
\label{equ-anomalous}
 g(\iwn)  = -\frac{1}{\beta}\sum_{\iwm} \Gamma^{pp}_c(\iwn,\iwm)
f(\iwm) 
\end{equation}
which is local in the DMFT. 
%%%
The anomalous Greens function $f$ can be replaced through the SCC
(\ref{eq-appB:off-media}) yielding the compact vector equation
\begin{equation}
\left[\mat{1} +  \frac{1}{\beta}\mat{\Gamma}_c \mat{A} \right]\ul{g}
= - \frac{1}{\beta} \mat{\Gamma}^{tt} \mat{A} \;  \mat{B}^{-1} \ul{f}
\komma
\end{equation}
where 
\begin{equation}
[\mat{A}]_{n,m} = \delta_{n,m} e(\iwn) h(\iwn)  
\end{equation}
and
\begin{equation}
[\mat{B}]_{n,m} = \delta_{n,m} \tilde e(\iwn) \tilde h(\iwn)
\punkt  
\end{equation}
Here, we introduced the short hand notation 
\begin{equation}
\ul{g} =
(\cdots , g(\iwn) ,\cdots )^T \punkt  
\end{equation}
We use 
the irreducible vertex  (\ref{eq:eff-pp-vertex-tt}) and note
that $\mat{\chi}_{loc} = -\mat{A}$ to obtain
\begin{equation}
\ul{g} = - \left[ \frac{1}{\beta} \mat{A}^{-1} \mat{\Pi}_{loc}
+\mat{1}\right]   \mat{B}^{-1} \ul{\tilde f}
\punkt
\end{equation}
It has been discussed in section \ref{sec:local-two-part-prop} that the local two
particle \gf\  $\mat{\Pi}_{loc}$ (\ref{eq:eq-ph-cc-propagator}) 
can be written as
\begin{equation}
 \mat{\Pi}_{loc} = -\beta \mat{A} +  \mat{B}\; \mat{T}_c \;  \mat{B}
\komma
\end{equation}
where $\mat{T}_c$ is the local two-particle $T$-matrix cumulant
\begin{equation}
\mat{T}_c = V^4
\left(\mat{\Pi}_{f}^{pp} - \beta \mat{F}
\right)
\komma
\end{equation}
and $\mat{\Pi}_{f}^{pp}$ the local two-particle f-\gf. Hereby,
$[\mat{F}]_{n,m} = F(\iwn) F(-\iwn)\delta_{n,m}$.
%%Hence, Eqn.~(\ref{eq-appB:anomalous-sigma}) reads
Hence, Eqn.~(\ref{equ-anomalous}) reads
\begin{equation}
\ul{g} = -  \frac{1}{\beta} \mat{A}^{-1} \;  \mat{B}\; \mat{T}_c
\ul{\tilde f} \punkt
\end{equation}
With the matrix equation for the vector \\
$\ul{T}_s = (\cdots ,
T_s(\iwn), \cdots)^T$ %%and $\ul{\tilde f}$
\begin{equation}
 \ul{T}_s = -V^4 \frac{1}{\beta}\mat{\Pi}_{f}^{pp}\ul{\tilde f}
\komma
\end{equation}
which can be derived via equation of motion or functional derivative,
and 
\begin{equation}
\label{eqn:e-tilde}
\frac{\tilde e}{e} = \frac{1}{1+\tilde e T_e}
\hspace*{10mm}
\mbox{and}
\hspace*{10mm}
\frac{\tilde h}{h} = \frac{1}{1+\tilde h T_h}
\komma
\end{equation}
we obtain
\begin{equation}
\label{eqn:g}
g =  \frac{1}{(1+\tilde e T_e)(1+\tilde h
T_h)}
\left(T_s - T_e T_h \tilde f
\right)
\end{equation}
for each frequency $\iwn$ which is just the off-diagonal matrix
element of the $2\times 2$ self-energy matrix $\mat{\Sigma}$ 
\begin{equation}
\mat{\Sigma} = \mat{T} \left[\mat{1} + \mat{\tilde G} \mat{T} \right]^{-1}
\end{equation}
in the limit $f,\tilde f \rightarrow 0$
where the one-particle $T$-matrix is given by
\begin{equation}
\mat{T} = \left( \begin{array}{cc}
           T_e & T_s \\
           T_s & T_h
          \end{array}
          \right)
\punkt
\label{eqn:b19}
\end{equation}
In Eqns.\ (\ref{eqn:e-tilde} -\ref{eqn:b19}), %% and (\ref{eqn:g}), 
the energy argument $\iwn$ was dropped for simplicity.
In general, $\mat{\Sigma}$ reads
\begin{eqnarray}
\label{eq-appB:sigma-sc}
\mat{\Sigma} & =  &
\frac{1}{A} 
\left(
\begin{array}{cc}
T_e +\tilde h(T_e T_h - T_s^2) & 
T_s - \tilde f (T_e T_h - T_s^2)\\
T_s - \tilde f (T_e T_h - T_s^2) &
T_h +\tilde e(T_e T_h - T_s^2)
\end{array}
\right)
\\
A & = & (1 +\tilde e T_e + \tilde f T_s)
(1 +\tilde h T_h + \tilde f T_s)
\non
&&
-(\tilde e T_s +\tilde f T_h)
(\tilde h T_s +\tilde f T_e)
\punkt
\end{eqnarray}

As mentioned above, $g(\q,\iwn)$
is generated by the par\-ticle-par\-ticle irreducible vertex
$\Gamma^{pp}_c(\iwn,\iwnn)$, Eqn.~(\ref{equ-anomalous})
\begin{equation}
\label{equ-anomalous-2}
 g(\q,\iwn)  = -\frac{1}{N\beta}\sum_{\k,\iwnn} \Gamma^{pp}_c(\iwn,\iwnn)
f_{\k,\q}(\iwnn) 
\end{equation}
which is local in the DMFT. On the other hand,  we can expand the
$\k$-dependent anomalous  
Green's function $f_{\k,\q}(\iwn)$  close to $T_c$ up to linear order 
\begin{equation}
\label{eqn:expansion}
f_{\k,\q}(\iwn) = -G_{\k+\q}(\iwn)G_{-\k}(-\iwn) \, g(\q,\iwn)
+ O(g^2)
%%\hspace{10mm}
\end{equation}
in the anomalous self-energy $g(\q,\iwn)$ for Cooper pairs with a
center of mass momentum $\q$. 
Inserting (\ref{eqn:expansion}) into (\ref{equ-anomalous-2}) and
identifying
\begin{equation}
\chi_{0}^{pp}(\iw{n};\ivn,\q \, ) = \frac{1}{N}\sum_{\k}
G_{\k+\q}(\iwn+\ivn)G_{-\k}(-\iwn)  
\end{equation}
yields
\begin{equation}
\label{equ:appB-linerarized-eliashberg-s}
 g(\q,\iwn)
 =  
\frac{1}{\beta}\sum_{\iwm} \Gamma^{pp}_c(\iwn,\iwm)\chi^{pp}_{0}(\q,0)
g(\q,\iwm)
\punkt
\end{equation}
If we set $\ul{g} \propto \ket{\mbox{eliash}} = \left(
  \mat{\chi}(\q=0,0)\right)^{-\frac{1}{2}}\ket{\lambda=1} $, where
$\ket{\lambda}$ was a solution of the eigenvalue equation
(\ref{eq:base-m}) $\mat{M}\ket{\lambda} = \lambda \ket{\lambda}$, 
the vector $\ket{\mbox{eliash}}$
is a solution of the linearized Eliashberg equation
(\ref{equ-linerarized-eliashberg-s}). Since $\ket{\lambda=1}$
corresponds to a diverging pair-susceptibility $\chi_P$
(\ref{eq:pair-sus-lambda}), a singularity in  $\chi_P$  is equivalent
to a solution of the  SCC (\ref{eq-appB:dmft-scc}) close to $T_c$ with
non-vanishing off-diagonal matrix elements.

%\bibliographystyle{prsty}
%\bibliography{references}

\end{fmffile}

\end{document}